\documentclass[12pt,a4paper,twoside,textlogo,CEJP]{cesj}
\usepackage{epsfig}
\usepackage[usenames]{color}
\def\d{\rm d}
\def\e{\rm e}
\def\k{\bf k}
\def\x{\bf x}
\def\v{\bf v}
\def\B{\bf B}

\issue{2(2)}
\year{2004}

\title{Propagation of Ultra-High Energy Cosmic Rays in Extragalactic Magnetic Fields}
\author{Tadeusz Wibig\inst{1,2} \email{E-mail: wibig@zpk.u.lodz.pl}}

\institute{Experimental Physics Department,\\ University of \L \'{o}d\'{z}\\
Pomorska 149/153, 90-236 \L \'{o}d\'{z}
\and
The Andrzej So\l tan Institute For Nuclear Studies,\\
Cosmic Ray Laboratory  \\Uniwersytecka 5, POB 447, 90-950 \L \'{o}d\'{z} 1, Poland}

\received{23 December 2003}
\accepted{21 February 2004}     %% These are for published papers.

\leftmark{T. Wibig}
\rightmark{T. Wibig}
\abstract{In this paper we will discuss the problem of Ultra High Energy Cosmic Rays (UHECR)
and show that the idea of a Single Source Model established by Erlykin and Wolfendale (1997)
to explain the features seen in cosmic ray energy spectra around the 10$^{15}$~eV region can
be successfully applied also for the much higher energies. The propagation of
UHECR (of energies higher than 10$^{19}$~eV) in extragalactic magnetic fields can
no longer be described as a random walk (diffusion) process and the transition
to rectilinear propagation gives a possible explanation for the so-called Greisen-Zatzepin-Kuzmin (GZK)
cut-off which still remains an open question after almost 40 years.
A transient ``single source'' located at a particular distance and producing UHECR
for a finite time is the proposed solution.}


\begin{document}

\firstpage{277}
%\linenumbers
\maketitle
\setcounter{page}{277}%

\section{Introduction}
The phenomenon known as cosmic rays, and particularly
the observed flux of particles of extremely high energies, is
a perfect example of the situation where the subject of study is ``one and only'' in nature --
{\it i.e.,} we only have one set of data.
Thus, in trying to explain it, one does not have to rely on the ``most probable''
or ``average'' solution.
The phenomenon as we see it, here and now, could be the
result of a particular chain of coincidences. If this chain is not
``very improbable,'' it may just be the right solution.

This concept was used
by Erlykin and Wolfendale a few years ago \cite{ew97}
in the Single Source Model (SSM) of CR origin. Originally it
was established to explain the shape of the
so-called ``knee'' in the CR energy spectrum seen in many experiments
over a period of almost 50 years.
Careful analysis of very accurate
data on Extensive Air Showers
collected by different experiments made in \cite{ew01b}
shows the existence of sharp structures around the
estimated primary CR particle energy of a few times 10$^{15}$~eV.
In subsequent
papers by Erlykin and Wolfendale \cite{ew01all},
it was shown
that the Single Source Model
could be used to explain a number of observed CR phenomena.
Here, we are going to follow
the SSM idea and go further up in energy to the very end of the
cosmic ray energy spectrum.

Many of the experimentally observed features in the UHECR domain ({\it e.g.,}
the ani\-so\-tro\-py studies in \cite{ww01}), confirm
that  we actually see there the vanishing Galactic component and the new
Extra-Galactic (EG) one which starts to dominate above an energy of $3 \times
10^{18}$~eV. The analysis of all available data made in \cite{sww}
shows that the EG component may start as power-law with an index of about 2,
and then, above about $10^{19}$~eV, continue with
observed index of $\sim 3$ up to the end
of measurements ({\it i.e.,} 10$^{20}$~eV or slightly higher).

The CR sources, especially for ultra high energies,
are unknown. Two general classes of
potential sources have been studied in the literature:
(i) astrophysical objects,
such as active galactic nuclei (AGN), quasars, and colliding
galaxies (see, {\it e.g.,} \cite{alda}), where the usual cosmic matter
constituents are accelerated to extremely high energies in
so-called ``bottom-up'' processes;
and (ii) some exotic ``top-down'' mechanisms such as
the decay of (super-heavy) dark matter particles, topological defects,
or monopoles (see, {\it e.g.,} \cite{bere}) .

The serious problem for ``bottom-up'' theories
is the UHECR general isotropy.  There is
no significant excess in any direction
to a potential source.

In recent work, evidence has been presented
that UHECR particles have a distribution of masses
\cite{ww99},
generating obvious difficulties for ``top-down'' ideas.
This finding is essential to the present work---
a significant
fraction of UHECR particles are multiply
charged (up to $Z=26$ in the case of iron), which makes
them more sensitive to extragalactic magnetic fields.

The ``average'' approach to UHECR spectrum
calculations, found already in the first
Greisen, Zatzepin, and Kuzmin \cite{gra,zk}
papers, is to assume that because we know
nothing about the sources, that for every point
in space and time the production of UHECR
is equally probable. The UHECR spectra shown in \cite{zk}, but also the frequently quoted
spectrum published by the AGASA group \cite{takeda}, were obtained assuming constant
and uniformly distributed UHECR source power in the whole Universe.
It gives a perfectly isotropic distribution of UHECR directions and a clear GZK cut off,
which is the consequence of interactions of UHE nucleons with the 3~K cosmological microwave
background photons.
However, in the real Universe
the distribution of matter is not exactly uniform.
Structures known as galaxy clusters exist,
and, if the UHECR sources are astrophysical,
they should follow the usual matter (galaxies)
distribution. Our Galaxy is within the Virgo cluster, about 15 Mpc from its
center, and it is obvious that particles of
energies above 10$^{19}$~eV, if created there,
should point more or less exactly to their sources.
Some enhancement is actually seen, but it is
statistically not very significant, and,
as will be discussed in this paper, far too small when compared with
expectations.

Assumptions about some particular, non-uniform distribution of UHECR sources in extragalactic space
have also been carefully studied recently, in Refs. \cite{isol,bart}. The UHECR spectrum and
small and large scale correlations (anisotropies) calculated there are significantly closer to the
measured cosmic ray features than in models with a uniform source distribution.

The present work goes, in some senses, a step further in this direction.
A single source is certainly far from isotropy, but here we also reject the assumption about its
constancy in time. This introduces an additional parameter---the dimension of time,
but at the expense of requiring an essentially new solution of the general anisotropy problem, as will be shown below.

For a continuous UHECR source, the very energetic particles should
propagate along (nearly) straight lines, reaching the observer after a time $\approx R/c$ and giving
evident directional correlation with some astrophysical objects, which is not the case in practice.
%Some claims of observing such kind of correlation appeared, but they are not conclusive, anyway.

We will discuss here the possibility that the UHECR sources are of transient nature---that they are in an
active state for some time, say 10$^{7-9}$~years (an interval so big that it
covers the collision time for galaxies passing each other, the estimated time of activity of AGN, etc.),
and then remain quiet. UHECR are assumed to be produced only in the active phase.
The idea is that the bulk of UHECR were produced by one
or a few sources located relatively nearby
(on the  extragalactic scale), but which are at present not active. This is a simple solution
of the isotropy problem. The very energetic particles traveling rectilinearly have passed Earth already,
and what we see now as the UHECR flux is only those particles which are deviated enough
by extragalactic magnetic fields to be delayed, relative to the light signal, by a substantial amount of time.

The only problem is to see if such a mechanism can really work---if the magnetic fields are strong enough
to curve the trajectories of particles of energies around 10$^{20}$~eV.

\section{Propagation of UHECR in the Intergalactic Magnetic Field}

The UHECR under consideration are electrically charged, so their propagation in intergalactic space is therefore
affected by the magnetic fields along their path. The intergalactic magnetic field strength is believed to be on the order of
10$^{-8}$--$10^{-9}$~G, and for the distances of interest of about 1--100 Mpc and particle energies above
10$^{18}$~eV, some deviations from rectilinear propagation are expected.

Experimental knowledge of large-scale magnetic fields is rather scarce (see, for example,
\cite{kron} and discussions given in \cite{hara} and \cite{ww2004}).
These fields will have both regular and random components. The former can be, in principle, a relic of distant
epochs (occasionally compressed and magnified or amplified by dynamo-like mechanisms). However, at present we
have no evidence of the existence of such, so we neglect it.

The irregular component is present in intergalactic space, as it is in our Galaxy (and others).
Its source can be
ionized plasma emitted by galaxies and clusters of galaxies, some of which will have come from supernova
remnants bursting out of the host galaxies. The escape of galactic cosmic rays into the intergalactic medium (IGM)
is a special case of
this ``process.'' Insofar as the energy density of cosmic rays in the IGM---coming from escape from
galaxies, is $\sim10^{-6}$~eV~cm$^{-3}$ (obtained by integrating the extragalactic flux of
cosmic rays), the corresponding magnetic energy density will give an rms field of
$\sim 3\times 10^{-9}$~G assuming equipartition. Another source of extragalactic magnetic field
is from active galactic nuclei and other near-cataclysmic events.
The magnetic
disturbances evolve in time in accordance with the conventional turbulence picture, transferring energy
consecutively down to smaller scales where the energy is finally dissipated.

There are various possibilities for the manner in which particles propagate through the IGM, but here we consider
just two: the cubic domain model and the Kolmogorov turbulence model.

We now examine how the particular random field structure influences UHECR propagation across large distances.

\subsection{Cubic domain model for the random magnetic field}

The transport of charged particles when well-known conditions are fulfilled can be described as a diffusion
process. The diffusion itself can be thought of as the limit of the constant step random walk process, this being
defined by one parameter only: the length of a single step.

\begin{figure}[ht]
\centerline{\epsfig{file=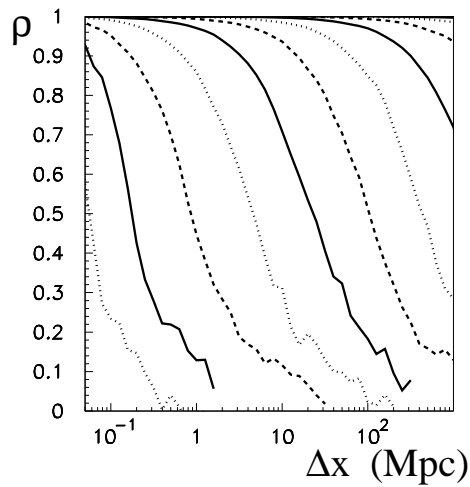,width=7cm}}
\caption{The velocity direction correlation coefficient
for the cubic domain model of the random magnetic field shown as a function of separation distance for different
proton energies. The solid lines represent energies of $10^{18}$, $10^{19}$, $10^{20}$, and $10^{21}$~eV (from the
least correlated to the one close to 1 for almost all distances), dashed and dotted 2 and 5 times these values,
respectively. \label{corrcub}}
\end{figure}

On the other hand, there is a limit to the large-scale random magnetic field arising from the results of Faraday
rotation measurements. It can be said that
\begin{equation}
\left\langle \B_{\Vert}\right\rangle \times \sqrt{\lambda_B} < 10^{-9} {\rm G}\:{\rm Mpc}^{1/2}~, \label{RM}
\end{equation}
where $\lambda_B$ is the magnetic field coherence length, which can be treated as the distance over which
the orientation of the magnetic field changes randomly. The simplest model of a chaotic magnetic field is just the
cubic domain model, in which the space is divided into equal cubic cells of size $\lambda_{\rm cell}$, the
field in each cell is equal to $\left\langle B\right\rangle$, and its orientation changes randomly from cell to
cell.

In such a picture, due to the fact that the cubic lattice orientation as a whole is obviously not fixed, the
effective coherence length $\lambda_B$ is defined precisely as
\begin{equation}
\int \B(\x) \cdot \B(0) \d l ~= ~ \left\langle \B^2 \right\rangle \lambda_B
\label{lc}
\end{equation}
(where the integration goes along the straight line over a distance much greater than any of the
regular component scales of $\B$). $\lambda_B$
is not exactly equal to $\lambda_{\rm cell}$, but the difference for our purposes (extragalactic UHECR
propagation) is not significant.

The magnetic field coherence length in the case of UHECR cannot be used as a random walk step size for the
propagation calculations. If the Larmor radius of a particle of charge $Z$ and energy $E$ is bigger than
$\lambda_B$, then after traversing the distance $\lambda_B$ the particle velocity still remembers (on average) its
initial direction.

To find out the random walk step length, we performed simulations of charged
particles in a magnetic cubic lattice
of size 0.1 Mpc with a random magnetic field of 10 nG. This is comparable with the Larmor radius
($\propto E/(ZB)$) for protons of energy $10^{18}$~eV.
The propagation coherence length $\lambda_c$ is defined, by
analogy with Eq.(\ref{lc}), as
\begin{equation}
\int
 \v(\x) \cdot
\v(0) \d l ~= ~ \left\langle {\v^2 }\right\rangle \lambda_c~. \label{lam}
\end{equation}
The integration is similar to that in Eq.(\ref{lc}).
It is a function of particle charge and energy.
%The scaling with $Z$ (if particles do not lose their energy during
%the propagation) is obvious, but for $E$ it is not the case.

To study this in more detail, we plot in Fig.\ref{corrcub} the correlation coefficient for the proton velocity
direction defined as
\begin{equation}
\rho(\Delta \x)~=~{ \left\langle \v (x) \cdot \v (x+\Delta x) \right\rangle
\over \left\langle \v ^2 \right\rangle } \label{corr}
\end{equation}
for different energies traversing our cubic magnetic domain space (where energy losses are neglected).

\subsection{A turbulent random magnetic field}

A more realistic picture of the intergalactic magnetic field uses the Fourier modes and their power spectrum
\begin{equation}
\B(x)~ =~ \int {\d ^3 k \over (2 \pi)^3 }\:\B(\k) \e ^{\imath(\k \cdot \x +\phi(\k ))}~, \label{fou1}
\end{equation}
where $\phi( \k )$ are random phases, and $2\pi/L_{\rm min}<k<2\pi/ L_{\rm max}$ with $L_{\rm min}$  and $L_{\rm
max}$ are the lower and upper limits of the magnetic field turbulence scales, respectively.

\begin{figure}
\centerline{\epsfig{file=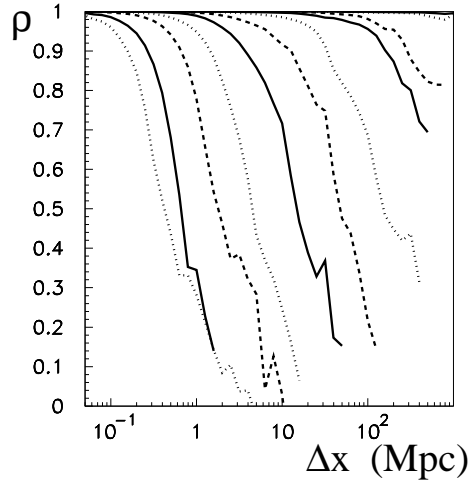,width=7cm}}
\caption{The velocity direction correlation coefficient for the
Kolmogorov turbulent magnetic field model. The key to the lines is as given
in Fig.\ref{corrcub}. \label{corrko}}
\end{figure}

In the present paper, a particular turbulent random field was realized by
replacing the integration in
Eq.(\ref{fou1}) by the sum of 1000 independent Fourier components, each with randomly chosen value of $k$ (limited
by $L_{\rm min}$ and $L_{\rm max}$) and random phase $\phi$. The sum was then normalized to yield the assumed
$\left\langle\left|\B(\x)\right|^2 \right\rangle$. In the calculations, we have used $\left\langle \left| B
\right|\right\rangle ~=~2 \times 10^{-9}$ G and $L_{\rm min}~=~0.01 \div 0.1$ Mpc and $L_{\rm max}~=~2$ Mpc.
Concerning the UHECR transport problem, the lower turbulence size limit is of no importance, and
the upper limit (in the reasonable range given above) has only a minor influence on the normalization of
$\left\langle \left| B \right|\right\rangle)$. The average value of $\B^2$ here is different from the one
assumed for the cubic cell model (as well as the scale of its irregularities),
but the propagation of charged particles just for such values is similar in both models,
as will be shown below.

The power spectrum $B^2(k)$ is proportional to the energy density contained in the $\k$ mode. For the power-law
turbulence spectrum,
%\begin{equation}B^2(k)~\sim~\langle \left|\B(\x)\right|^2 \rangle \: k^{-n}
%
%\end{equation}
\begin{equation}
B^2(k)~\sim~ \left\langle \left|\B(\x)\right|^2 \right\rangle\: k^{-n}
%{ (n-1) \left(2\pi/L_{\rm max}\right)^{n-1}
%\over 1\:-\:\left(L_{\rm min}/L_{\rm max}\right)^{n-1} }
~. \label{powerlaw}
\end{equation}
For the general case of Kolmogorov turbulence, the index $n$ is equal to $5/3$.
The magnetic field coherence
length for this case can be calculated analytically
\cite{hara}
%(Harari et~al., 2002, Bossa et~al., 2003)
and is equal to $L_{\rm max}/5$ for small
$L_{\rm min}/L_{\rm max}$.

For proton propagation in the Kolmogorov turbulent magnetic field, the correlation coefficient for velocity
direction $\rho$ given by Eq.(\ref{corr}) has been calculated and the results are given in Fig.\ref{corrko}.

\subsection{Comparison of particle propagation in random magnetic field models}
\begin{figure}[ht]
\centerline{\epsfig{file=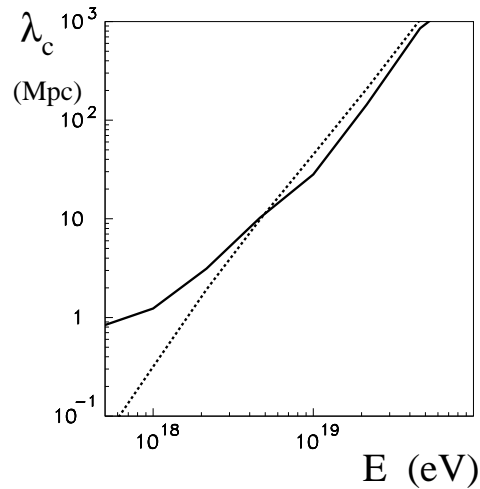,width=7cm}}
\caption{The propagation coherence length for the Kolmogorov
turbulent magnetic field model (solid line) and the cubic domain model (dashed line), for which
$\lambda_c\propto E^2$. \label{lambdac}}
\end{figure}

The propagation coherence length $\lambda_c$ defined in Eq.(\ref{lam}) for the turbulent medium in comparison with
the one for the cubic domain model is shown in Fig.\ref{lambdac}.

\begin{figure}[ht]
\centerline{\epsfig{file=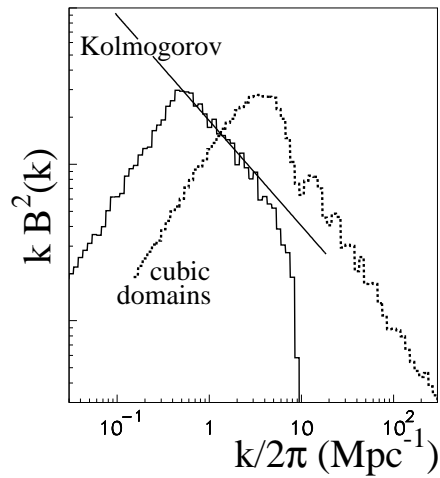,width=7cm}}%\epsfig{file=corrkd.eps,width=7cm}}
\caption{The powers of Fourier components calculated for the Kolmogorov
turbulent magnetic field model ($L_{\rm min}$
and $L_{\rm max}$ are 0.1 and 2~Mpc)
and the cubic domain model (with a cell size equal to 0.1 Mpc);
the straight line represents $k^{-5/3+1}$ dependence.
At right, the magnetic field correlation coefficient for each
model.
\label{fou2}}
\end{figure}

\begin{figure}[ht]
\centerline{\epsfig{file=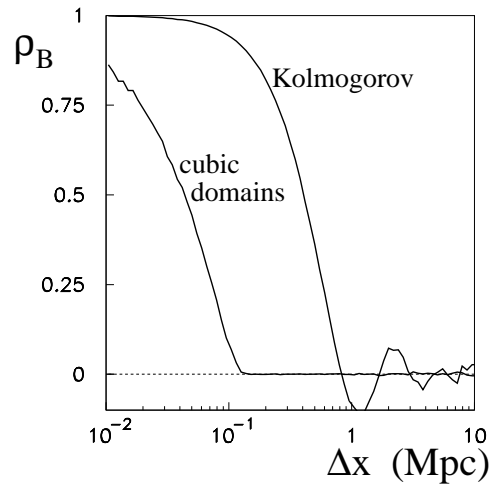,width=7cm}}
\caption{The magnetic field correlation coefficients for the
Kolmogorov turbulent magnetic field model and the cubic domain model.
(parameters as in Fig.\ref{fou2}).
\label{Bcorr}}
\end{figure}

It can be seen from all the figures that the transport of charged particles in the two types of random magnetic
field model should be very similar, in spite of the fact that the detailed structure of the field is so very
different. Not only is the average magnetic field strength $\left\langle \left| B \right|\right\rangle$ different
($10 \times 10^{-9}$ G for cubic domains and $2 \times 10^{-9}$ G for a Kolmogorov turbulent medium), but
the spectrum of the field is different. The spectrum $\left\langle
B^2(k)\right\rangle$ calculated by Fourier decomposition of generated chaotic fields in each model, and the
respective magnetic field correlation coefficients, are shown in Figs. \ref{fou2} and \ref{Bcorr}.
%\begin{figure}[ht]
%\centerline{\epsfig{file=btkd.eps,width=7cm}\epsfig{file=corrkd.eps,width=7cm}}
%\caption{(left): The power of Fourier components calculated for the Kolmogorov
%turbulent magnetic field model ($L_{\rm min}$and $L_{\rm max}$ are 0.1 and 2~Mpc)
%and for the cubic domain model (with a cell size equal to 0.1 Mpc). The straight line
%represent the $k^{-5/3+1}$ dependence.
%(right): The magnetic field correlation coefficient for both analyzed models.
%\label{fou2}}
%\end{figure}

\begin{figure}
\centerline{\epsfig{file=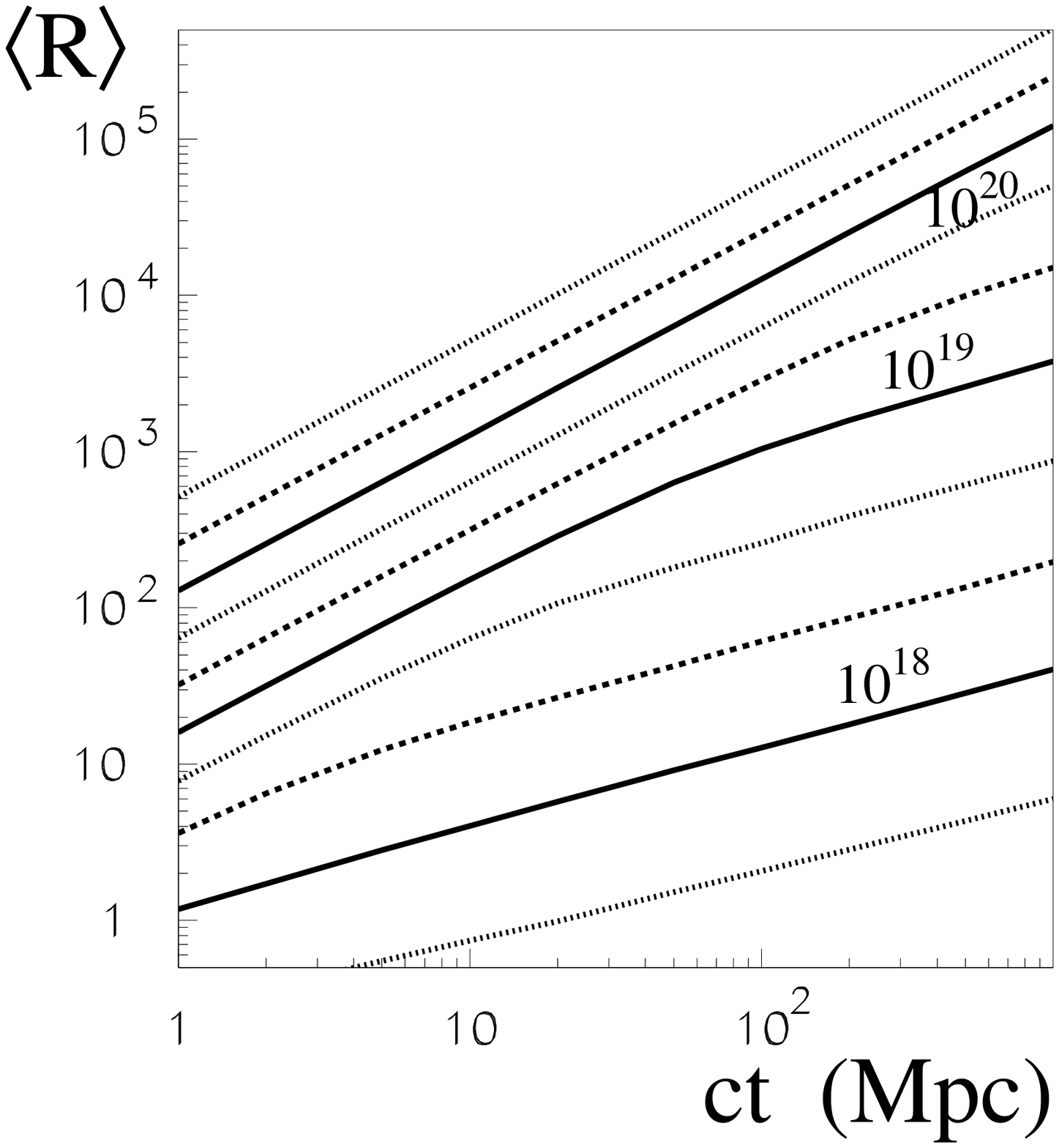,width=7cm}
\epsfig{file=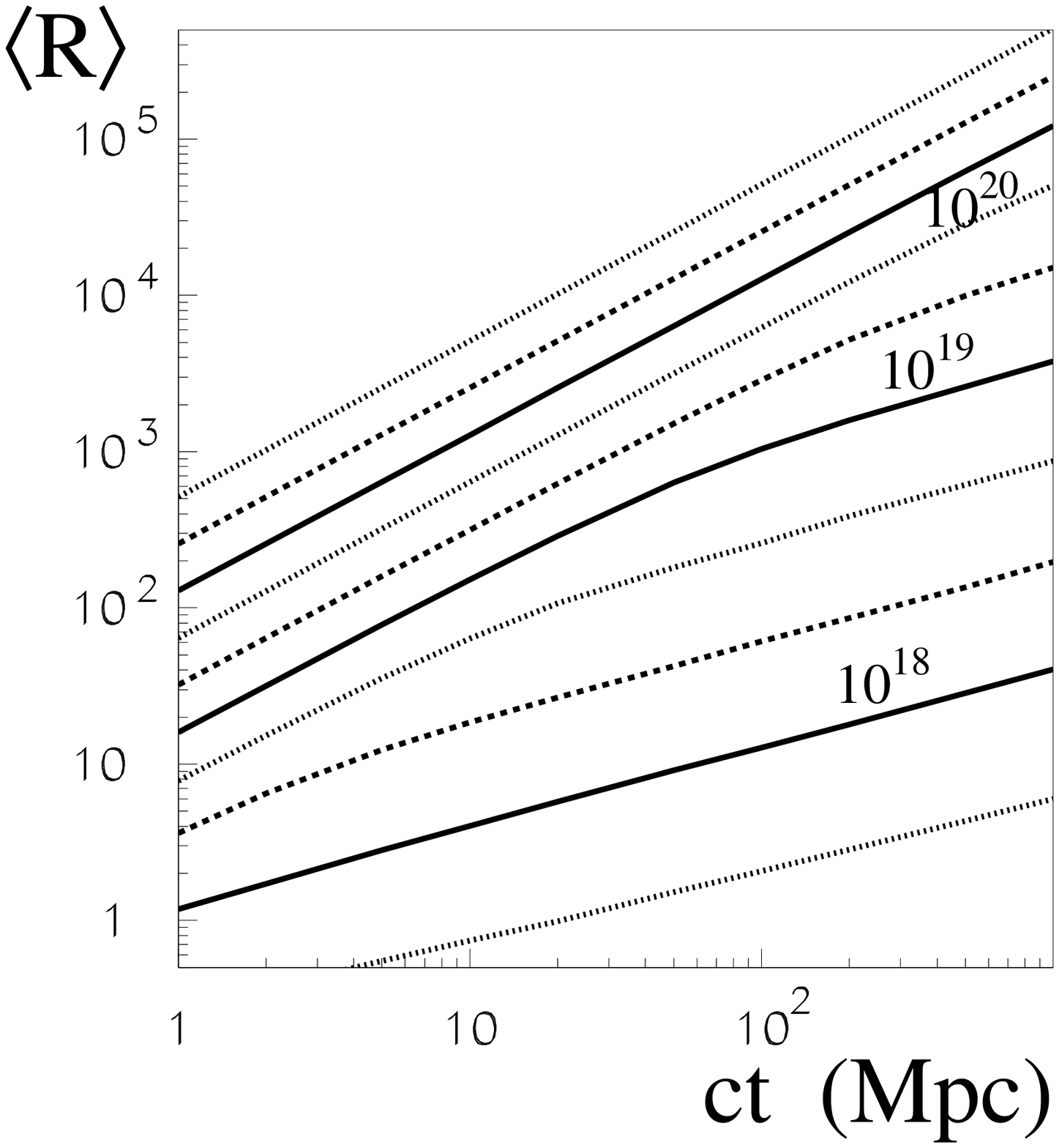,width=7cm}}
\caption{Mean distance traveled
by protons during time $t$ for the Kolmogorov turbulent magnetic field model (left) and for the cubic domain
model (right). A vertical shift was introduced to separate the lines for each particle energy. Parameters: $L_{\rm
min}=0.1$ Mpc and $L_{\rm max}=2$~Mpc; $\left\langle B^2\right\rangle=2\times10^{-9}$~G for the Kolmogorov
turbulent medium, and $\left\langle B^2\right\rangle=10^{-8}$~G and $\lambda_{\rm cell}=0.1$~Mpc for the cubic
domain model. \label{rvst}}
\end{figure}

\begin{figure}
%\vspace{-.5cm}
\centerline{
\epsfig{file=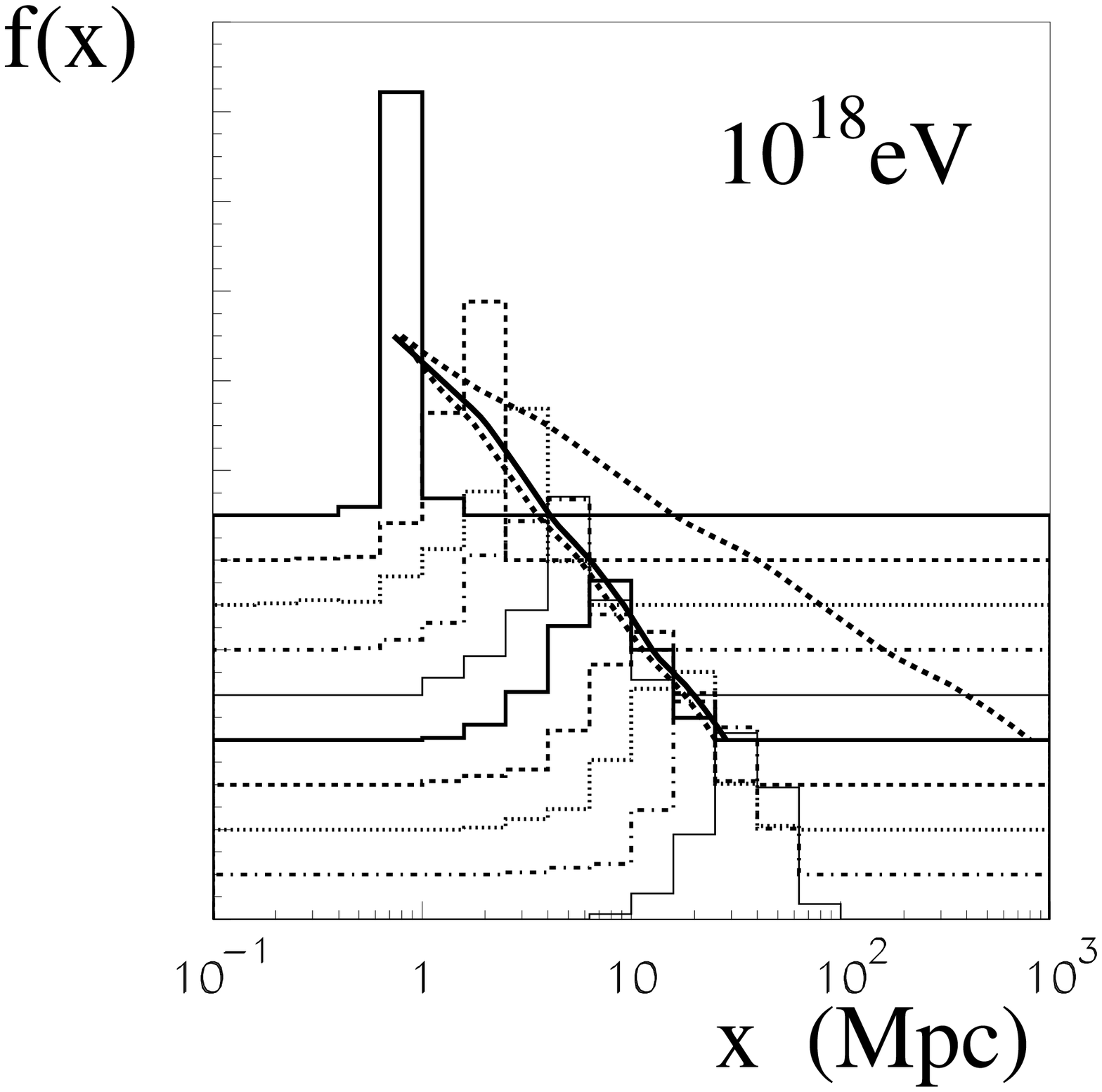,width=7.cm}
\epsfig{file=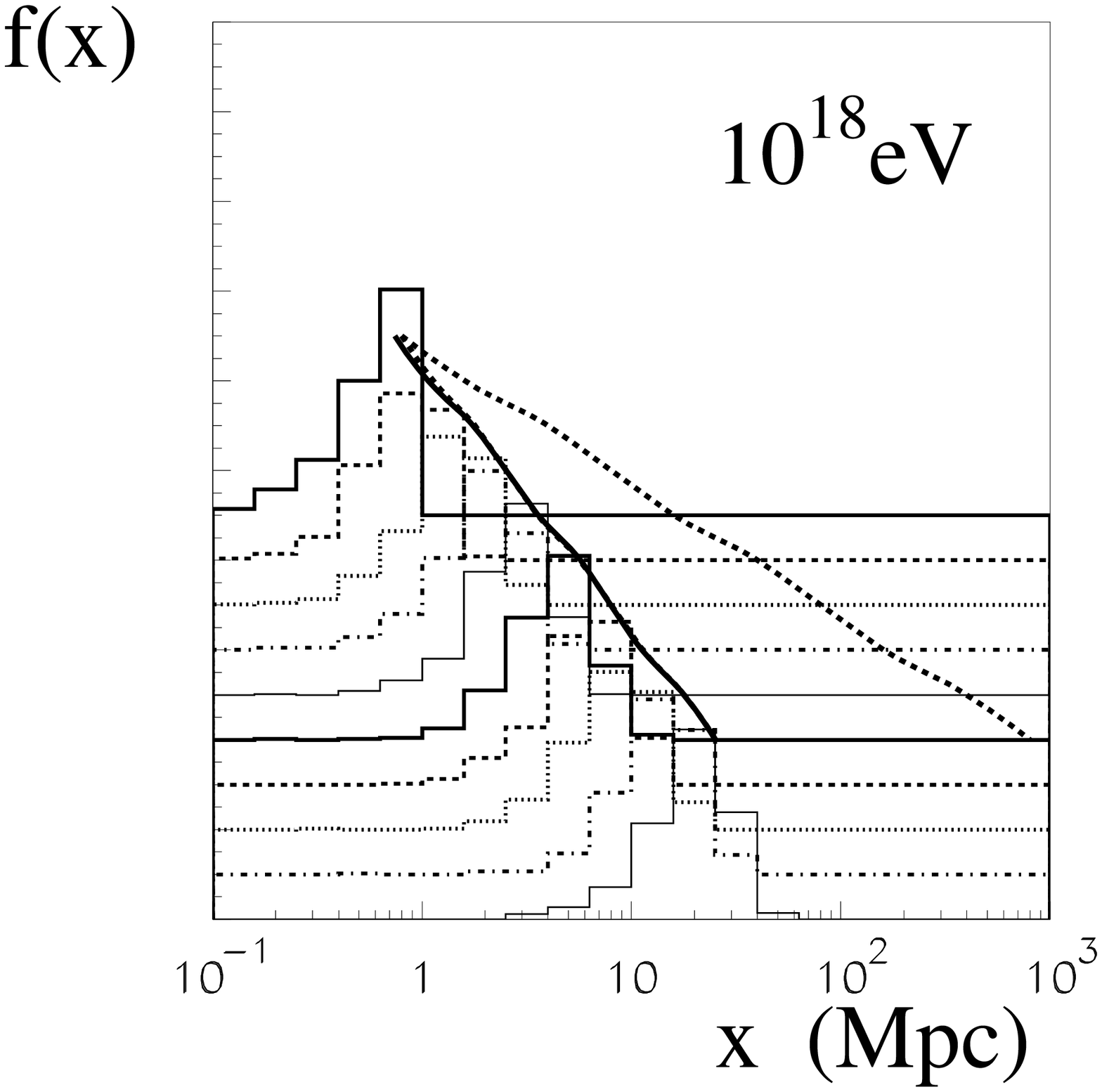,width=7.cm}
}

%\vspace{-.5cm}
\centerline{
\epsfig{file=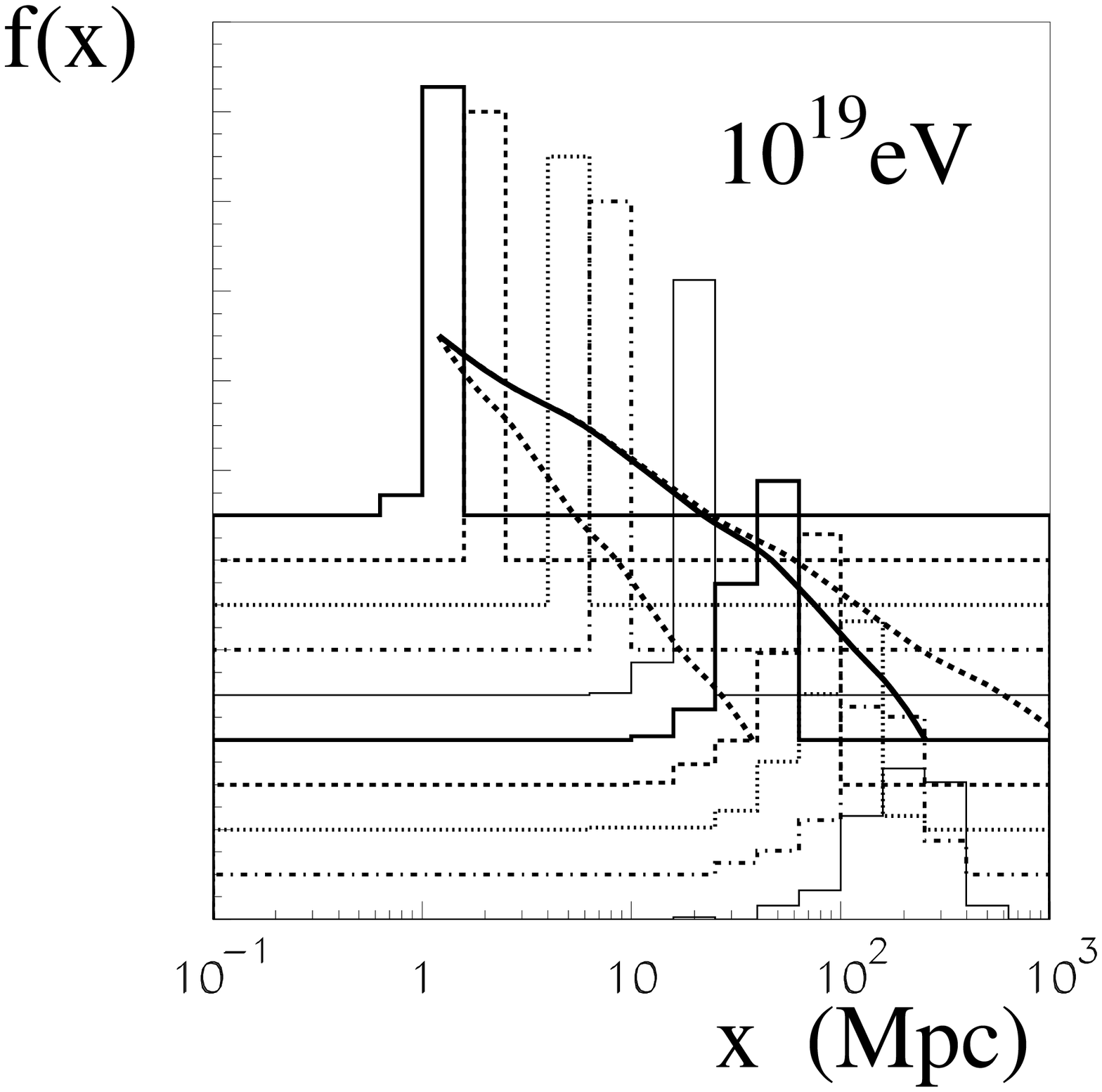,width=7.cm}
\epsfig{file=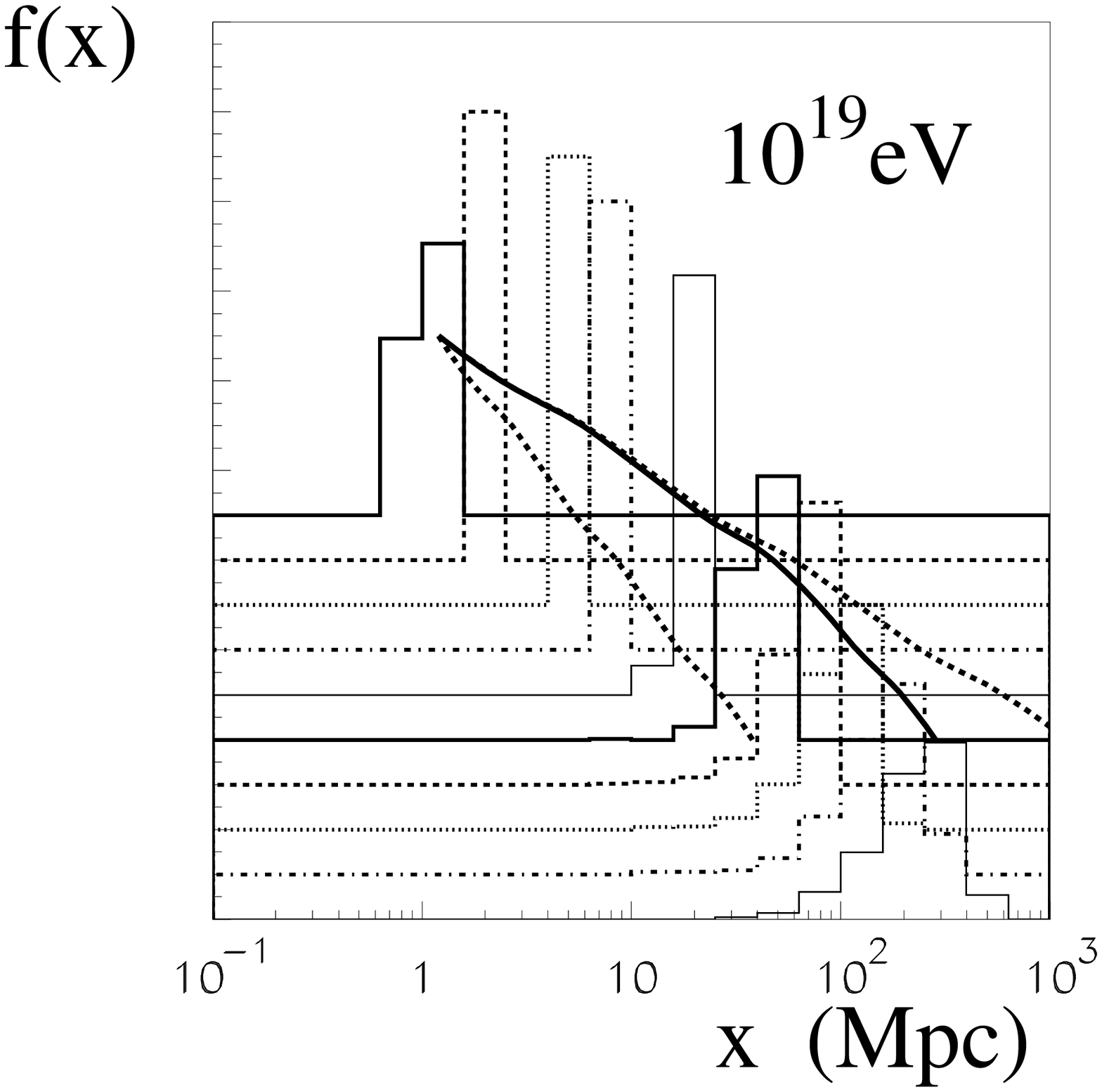,width=7.cm}
}

%\vspace{-.5cm}
\centerline{
\epsfig{file=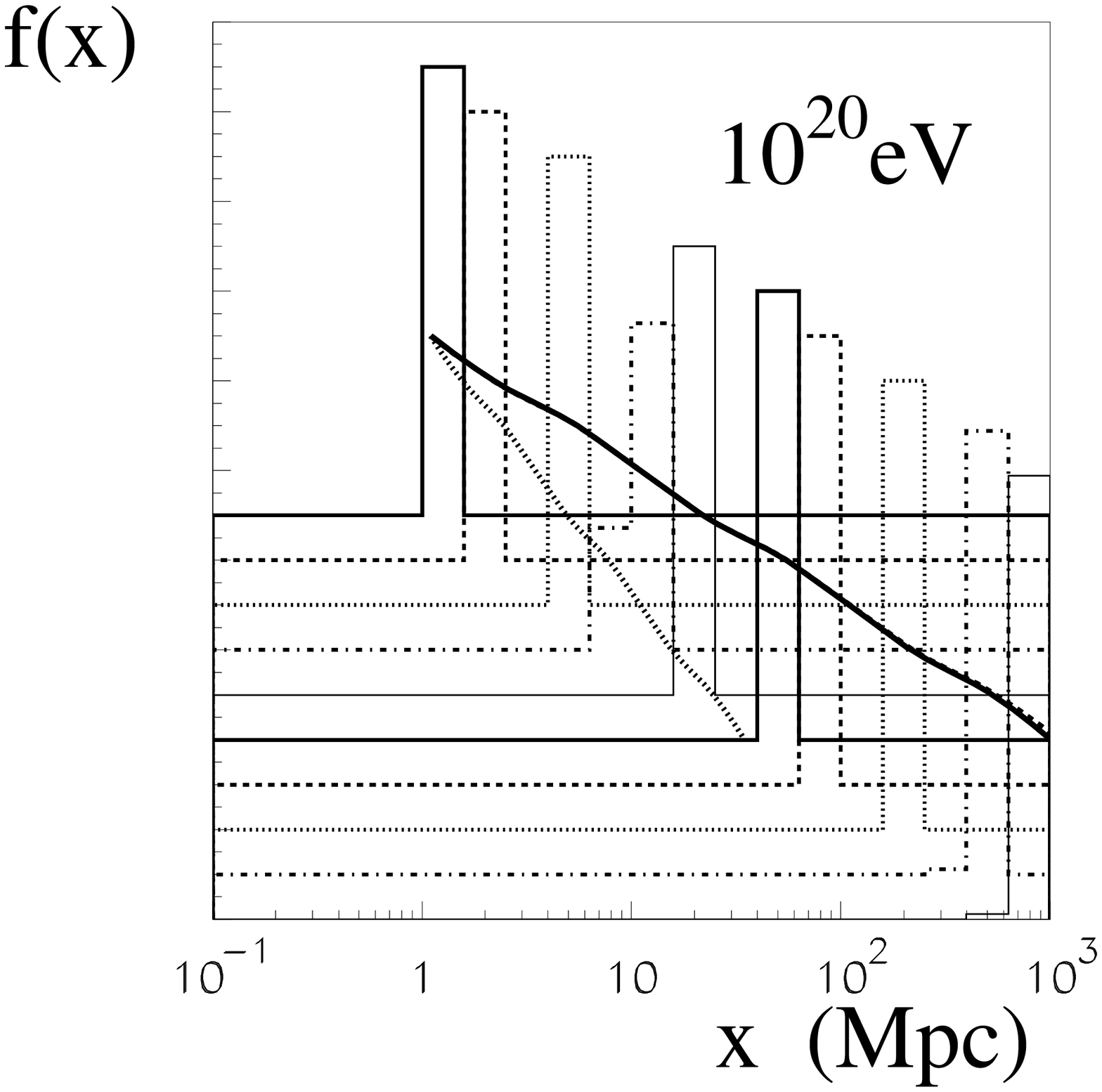,width=7.cm}
\epsfig{file=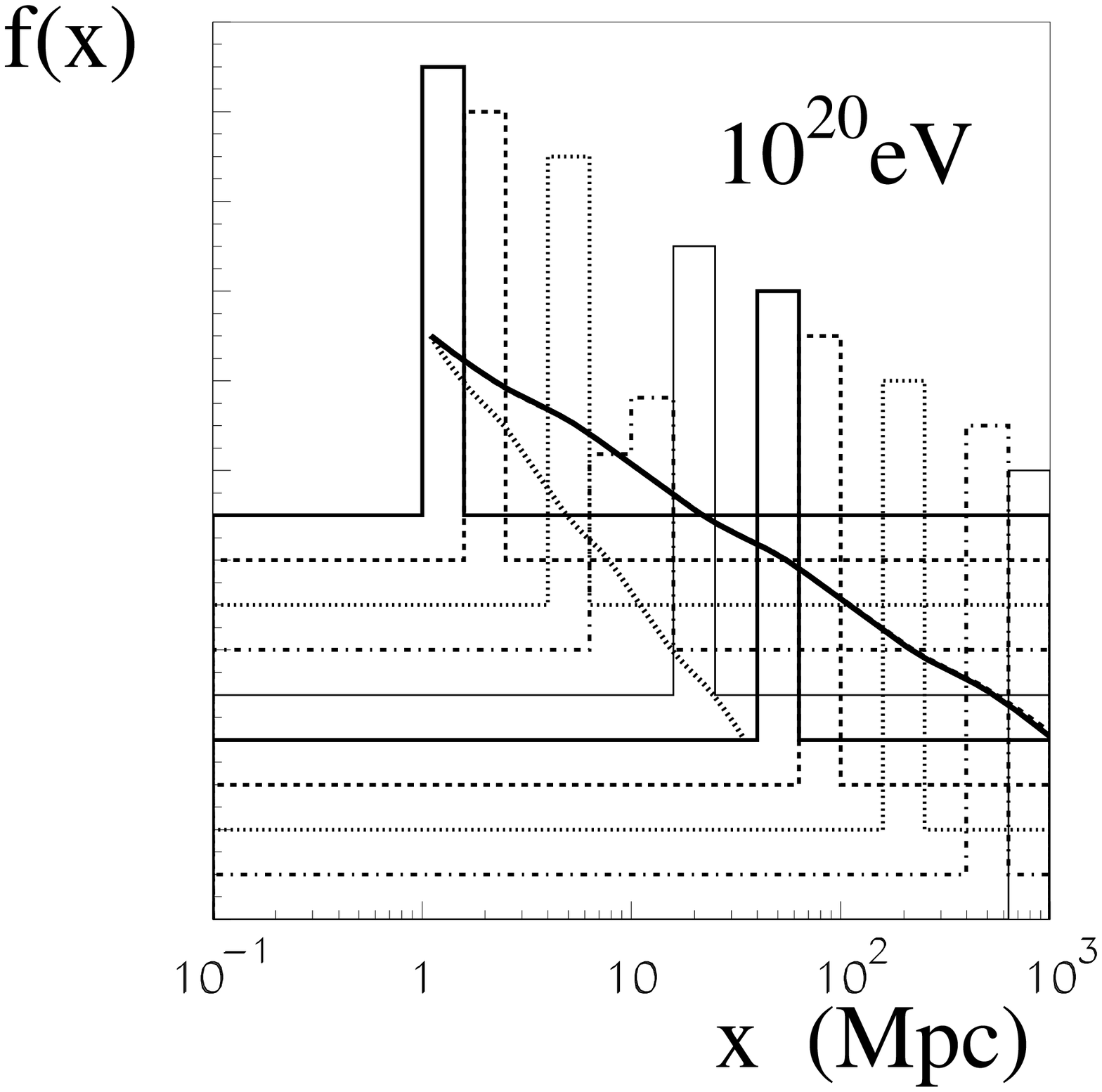,width=7.cm}
}
\caption{
Distributions of distance from source at different times of propagation
$ct=1,\:2 \:,5 \:,10\:,20 \:,50 \:,100 \:,200 \:,500 \:,$ and 1000~Mpc
(from top to bottom---the distributions are shifted
vertically). Inclined lines shows the positions
of the average values: dashed for the rectilinear
propagation ($R=ct$),
dotted for diffusion
($R \sim \sqrt{t}$),
and solid for the actual average as in
Fig.\ref{rvst}. The figures to the left are for the
cubic domain model, and those to the right for the Kolmogorov turbulent random magnetic field.
\label{rdst}}
\end{figure}

For a given energy, if the observer is located at a distance bigger than $\lambda_c$, as shown in
Fig.\ref{lambdac}, the propagation is diffusive.
The deviation from rectilinear propagation starts around $\lambda_c$.
This can be seen in Figs. \ref{rvst} and \ref{rdst}, where the mean
distance reached by the particle as a
function of time is shown, and where the
distance distributions are given.
For  rectilinear propagation, the respective line slope is approximately unity (on a log$\times$log) plot; when the
diffusion starts to dominate, the slope changes from $1$ to $1/2$. It is seen that particles with energies of
$10^{18}$~eV diffuse while those with energies of 10$^{20}$~eV propagate along (almost) straight lines, through distances of
the order of Gpc.

\section{Energy loss processes}
\begin{figure}
\centerline{\epsfig{file=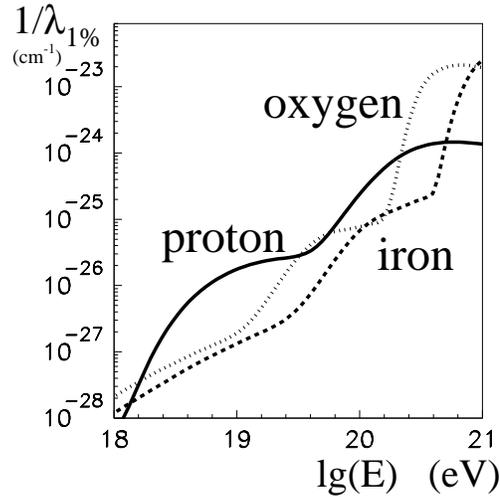,width=7cm}}
\caption{Inverse mean length for 1\%
 energy loss for iron, nitrogen, and protons
 on extragalactic background photons.
\label{1proc}}
\end{figure}

The UHECR domain is quite rich in physical processes involving energy losses. Starting with protons of
relatively low energies, about 10$^{18}$~eV,  $e^+e^-$ pair production on the cosmic microwave background (CMB)
photons starts to play a role, which reaches maximal importance slightly below 10$^{19}$~eV. The main GZK
process of energy loss is due to $\Delta$ resonance excitation (and its subsequent decay, dissipating energy,
eventually to low energy $\gamma$s) on CMB photons. The energy losses of heavier nuclei relative to
electron-positron pair creation are $Z^2$ stronger, but, due to the different rest mass and therefore different Lorentz
factor, the respective total nucleus energy should be $A$ times higher than that for protons. The same scaling in
energy ought to be applied for $\Delta$ resonance creation (but without the $Z^2$ enhancement). This
makes the GZK mechanism for heavy nuclei not as important.
The dominating process for nuclei is photo-disintegration on background photons. The
significant rise in fragmentation cross section just at the energies of our present interest is due to the
existence of giant dipole resonance.  This excitation energy is close to 20 MeV for (almost) all interesting
heavy nuclei. This is about one order of magnitude below the $\Delta$ resonance excitation energy, and thus, if
only the collisions  with CMB photons are considered, the threshold energy for nuclear disintegration is of order
$A/10$ higher than the proton GZK cut-off energy. A review of the whole situation is presented in
Fig.\ref{1proc} .

To compare the $e^+e^-$ pair production, $\Delta$ resonance creation by nucleons, and disintegration of
heavy nuclei, the cross
sections have to be convoluted, not only with the photon energy spectrum, but also with the inelasticity of the
respective process. In Fig.\ref{1proc}, the inverse average length for 1\% energy loss is shown.
It is different from the commonly used
$\left({\rm d}\ln(E)/{\rm d}x \right)^{-1}$ describing the
average length for losing a ($1-{\rm e}^{-1}\approx 63$\%) fraction of particle energy. But
it is actually more illustrative specifically for cases where cross sections change substantially
with the energy (by more than a decade for nuclei above 10$^{20}$~eV when the energy changes
by e$^{-1}$).  Anyway, the difference between our $\lambda_{1\%}$ and
$\left({\rm d}\ln(E)/{\rm d}x \right)^{-1}$ is in the constant factor $\approx 0.63/0.01$ by which
the vertical scale in Fig.\ref{1proc} should be multiplied to match the convention.

The CMB is assumed to be of temperature 2.7 K
and for higher energy photons we take the spectrum obtained in
\cite{malk}%Malkan and Stecker, 1998
(the one labeled there the
``best estimate'' intergalactic IRB).

Simulations of particle transport in both magnetic field models described above
were performed assuming that their source emits protons and composite nuclei
each time in newly generated field realizations. The particle trajectory was calculated
in small steps of 3~kpc (3~kpc/$c$ in time intervals). In each step, the field was assumed constant
and the trajectory was calculated analytically, giving position and velocity direction for the
traced particle after certain short time intervals.

Continuous energy losses were taken into account by diminishing the particle energy after each step.
Abrupt losses of particle energy due to photo-pion production and spallation in the case of nuclei were included by generating in each step
the actual
interaction lengths for reactions with one or two nucleons released (n, p, 2n, pn, 2p separately)
according to cross sections given in Ref.\cite{stesal}.
%Stecker and Salamon M.H. (1999).
If the shortest of these interaction lengths was within the spatial step length, the length of the step was reduced
to this value and the actual position of the interaction point (and particle direction
there) was calculated. In the case of photo-pion reaction, the average energy loss
due to $\Delta$ resonance decay was subtracted. For nuclei, all the secondary products
were included in the memory, to be propagated along with the initial nuclei until they were
eventually lost after reaching the overall energy threshold, or until the propagation
time limit (3~Gpc/c for the present calculations) was reached.

During propagation, particles were recorded each time they were within a
spherical shell of thickness of 100~kpc and radius 3, 5, 7, 10, 15,... Mpc around the
source. Their direction, energy, type (mass number), and time since emission
(or since the emission of their initial progenitor) were later used to obtain the
distributions of interest.

The initial energy spectrum was sampled in very short intervals in logarithmic scale and
integrated, weighting events by the power-law emission spectrum (with a differential
index of 2.1 for this paper).

\section{Small scale clustering of UHECR}
The UHECR, if they come from a relatively close source, are expected to
be directionally correlated. Their arrival directions
could point to the particular source in the sky.
Several attempts have been made to verify this hypothesis,
but all are based on limited statistics, and their
significance has been limited.

We present here some
results concerning the existence of small scale
clusters relevant to the subject of the present paper.
Our analysis is similar to the one in
\cite{clusagasa} based on the whole available Northern hemisphere
data on cosmic ray events of energies
above $4\times 10^{19}$~eV. The data consist of 113 events from AGASA,
Haverah Park, Yakutsk,
and Volcano Ranch experiments (19 of them with energies greater then $10^{20}$~eV).
We used a technique which was
developed in searches for correlations among particles created in
high energy accelerator experiments. Factorial moments in integral form are
the best tools to be used for our purpose.
Precisely, we used the so-called ``star integral'' method for factorial moment calculations
discussed extensively in \cite{lipa} and defined as
\begin{equation}
F_k(\Delta)~
=~%{1 \over  \left[\int_{\delta} \:\rho_1(y)\right]^k}~
{
\int \:
\rho_k(y_1,\:y_2,\ldots,\:y_k)\:\Theta_{12}\:\Theta_{13}\dots \Theta_{1k}\:
{\rm d} y_1\: {\rm d} y_2\: \ldots\:{\rm d} y_k
\over
\int \:
\rho_1(y_1)\:\rho_1(y_2)\:\dots\rho_1(y_k)\:\Theta_{12}\:\Theta_{13}\dots \Theta_{1k}\:
{\rm d} y_1\: {\rm d} y_2\: \ldots\:{\rm d} y_k
}~,
\end{equation}
where
$\rho_k(y_1,\:y_2,\ldots,\:y_k)$ is the $k$-dimensional probability density, and $\Theta_{ij}$
are equal to Heaviside step functions with argument $(\Delta - \|y_i,\:y_j\| )$:
\begin{equation}
\Theta_{ij}~=~\left\{
\begin{array}{lcl}
1& {\rm if}& \|y_i,\:y_j\| \le \Delta \\
0& {\rm if}& \|y_i,\:y_j\| > \Delta \\
\end{array}
\right.
\label{fact}
\end{equation}
where $\|y_i,\:y_j\|$ is the distance between two points defined in our case as the angle between
directions of UHECR events. The interpretation of factorial moments in integral form, thanks to the
$\Theta_{ij}$ functions, is clear.  The factorial moment gives the number of groups of events (doublets, triplets etc.) where the relative distances
within each group are smaller than $\Delta$ in the data sample analyzed, normalized by the
number of such groups calculated for the sample with the same marginal distribution for all
$y_i$ variables and lack of any correlation among any of them, {\it i.e.,} where
$\rho_{k\:{\rm norm}}(y_1,\:y_2,\ldots,\:y_k)=\rho_1(y_1)\:\rho_1(y_2)\:\dots\rho_1(y_k)$.

The normalization
factor can be obtained using the ``event mixing'' method, but
in general, factorial moments can be used for comparison of
observations with any model for the background.

Due to the small statistics, only the first two orders
could be studied with some confidence. The factorial moments are
related to integral cumulant moments $K$:
\begin{equation}
K_2~=~F_2\:-\:1
~,~~~~~
K_3~=~F_3\:-\:3\:F_2\:+\:2
\end{equation}
which represent genuine correlation of the given order present in the
sample analyzed.

\begin{figure}[ht]
\centerline{
\psfig{file=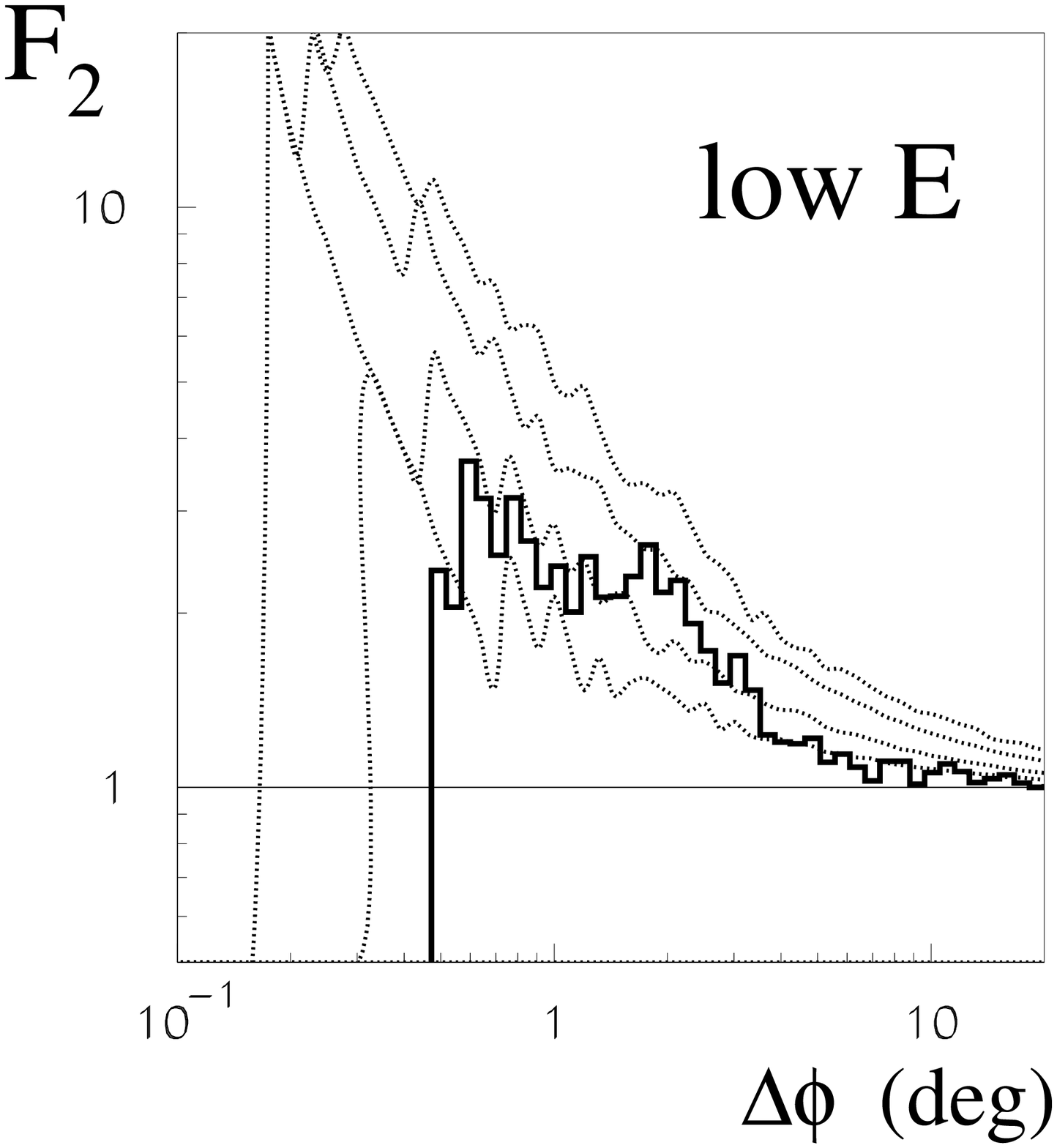,width=7cm}
\psfig{file=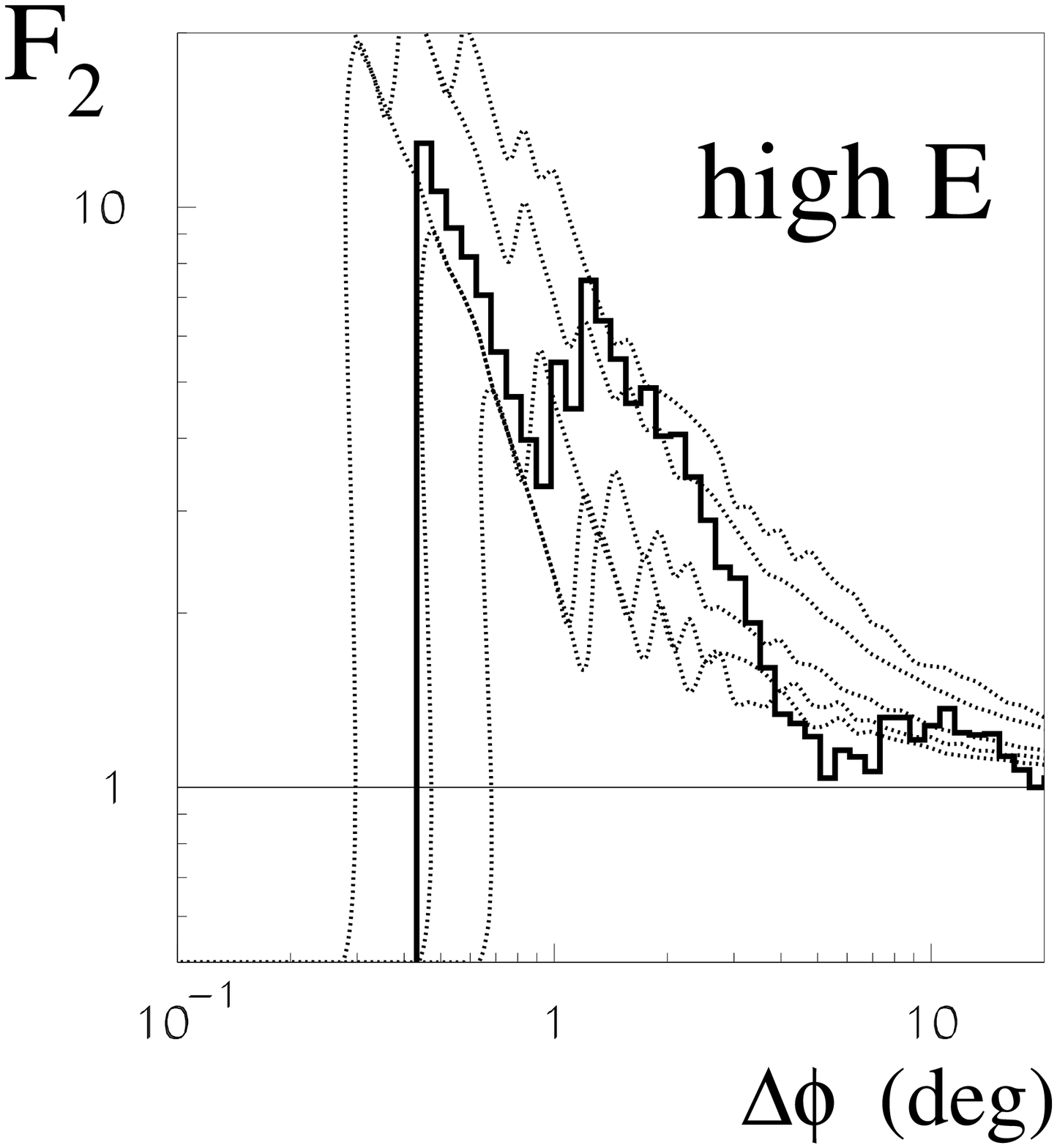,width=7cm}}
 \caption{Second factorial moment calculated for all event sample (``low E''---left),
and for pairs for which at last one event in the pair is of
energy greater than $10^{20}$~eV (``high E''---right).
The result of the data analysis is shown by the solid histogram.
Dotted lines represent the $90\%$, $95\%$, $99\%$, and $99.9\%$
confidence limits, respectively.
\label{f2k2}}
\end{figure}

In Fig.~\ref{f2k2}, results concerning two-point correlations are shown.
The observations are represented in the figures as the thick solid histogram.
To see the significance of the observed correlation,
upper limits can be calculated exactly using the Monte Carlo method by generating hundreds of thousands of
times the
uncorrelated ``mixed events'' pools and counting fractions of events exceeding each value of $\delta$.
The limits are shown as dotted histograms for confidence levels
of 90\%, 95\%, 99\%, and 99.9\%.

The analysis has been performed for two event samples.
In the first data sample (labeled as ``low E'' in the figures), all
events with energy of more than $4 \times 10^{19}$~eV were used and for
the second (``high E''), it was required that at least one event in the doublet or triplet
had to be of energy greater than 10$^{20}$~eV.
Such a division gives the possibility of checking if the correlation
is indeed increasing with the particle energy, as one would expect.

\begin{figure}[ht]
\centerline{
\psfig{file=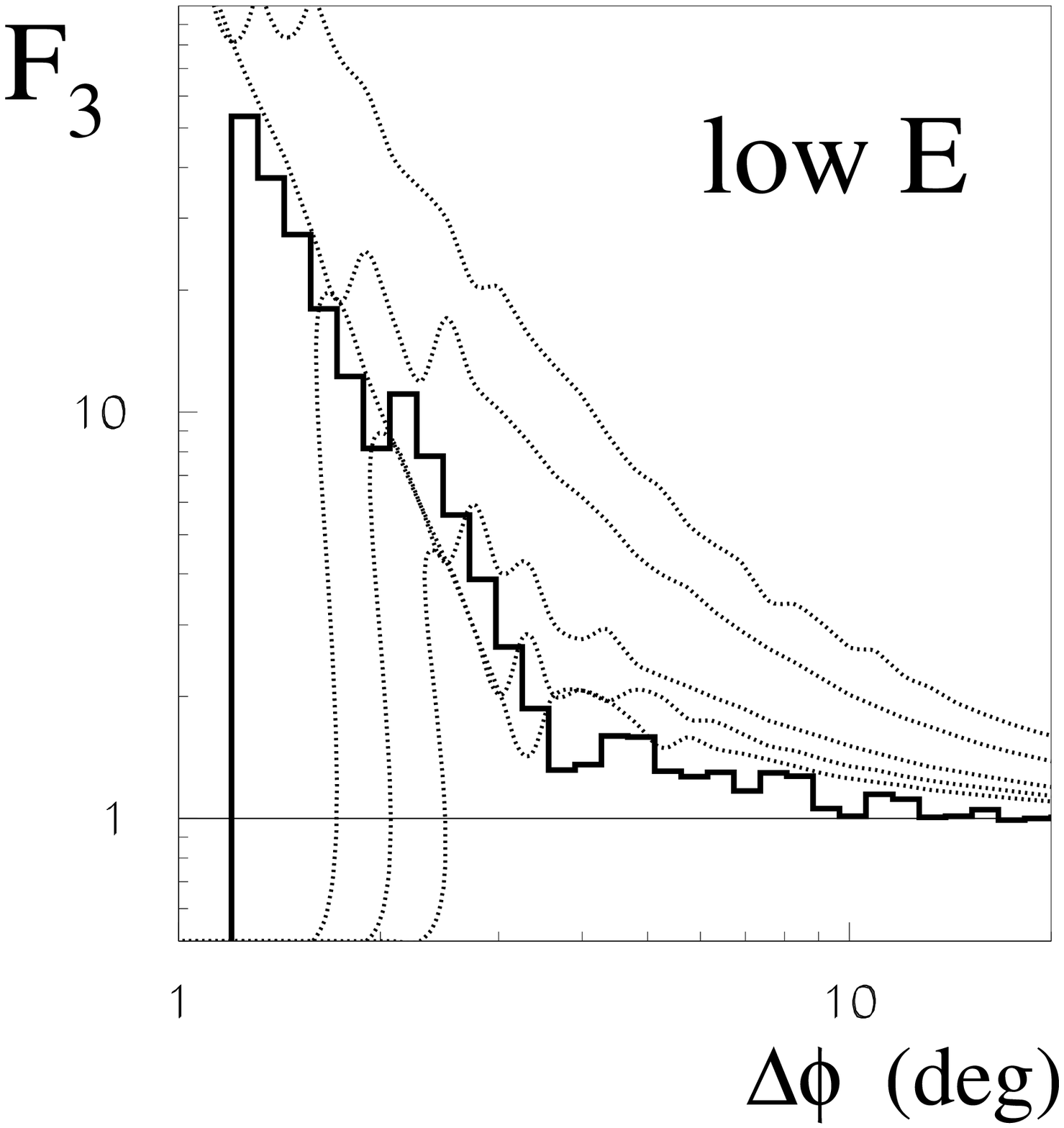,width=7cm}
\psfig{file=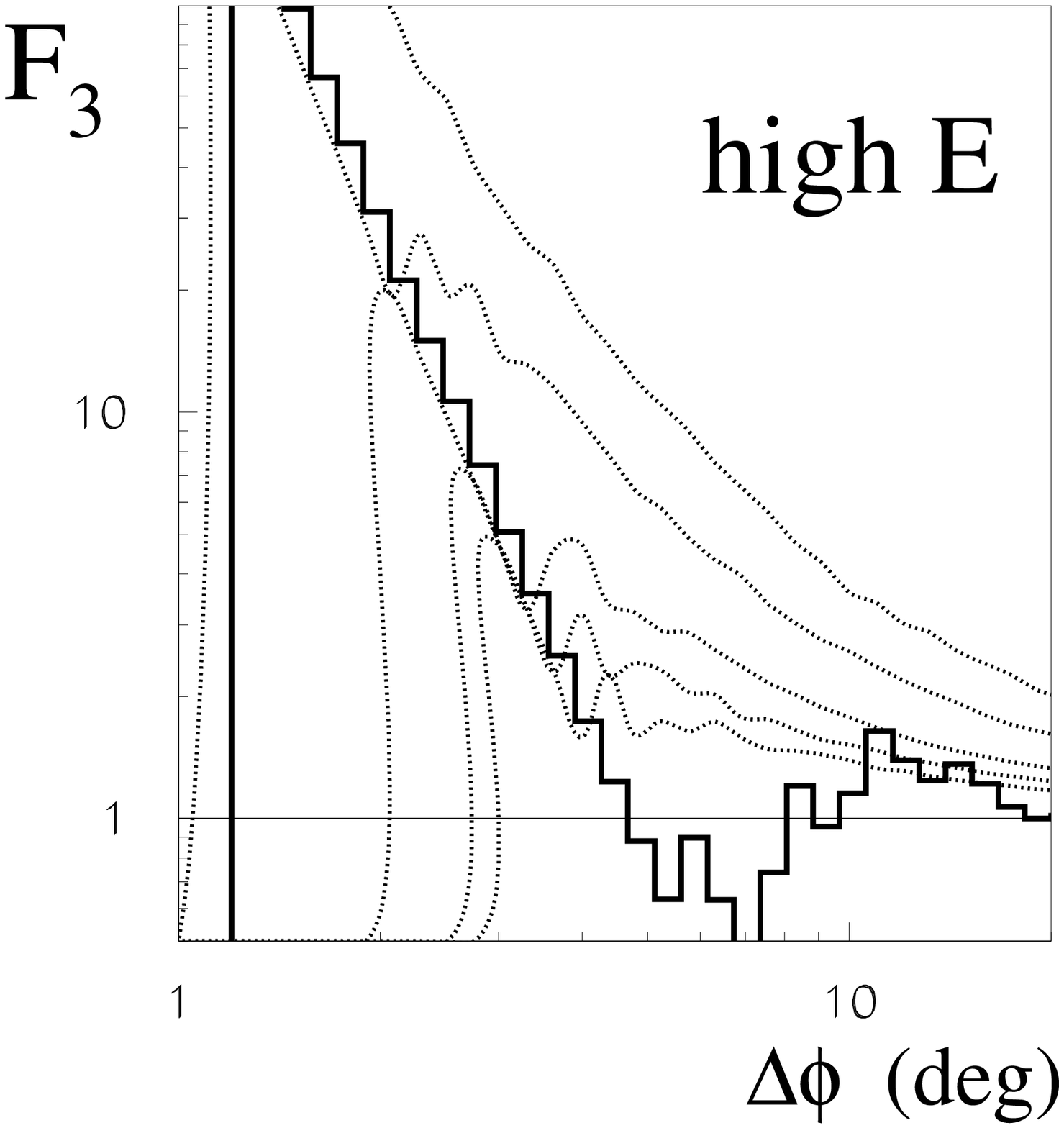,width=7cm}}
\centerline{
\psfig{file=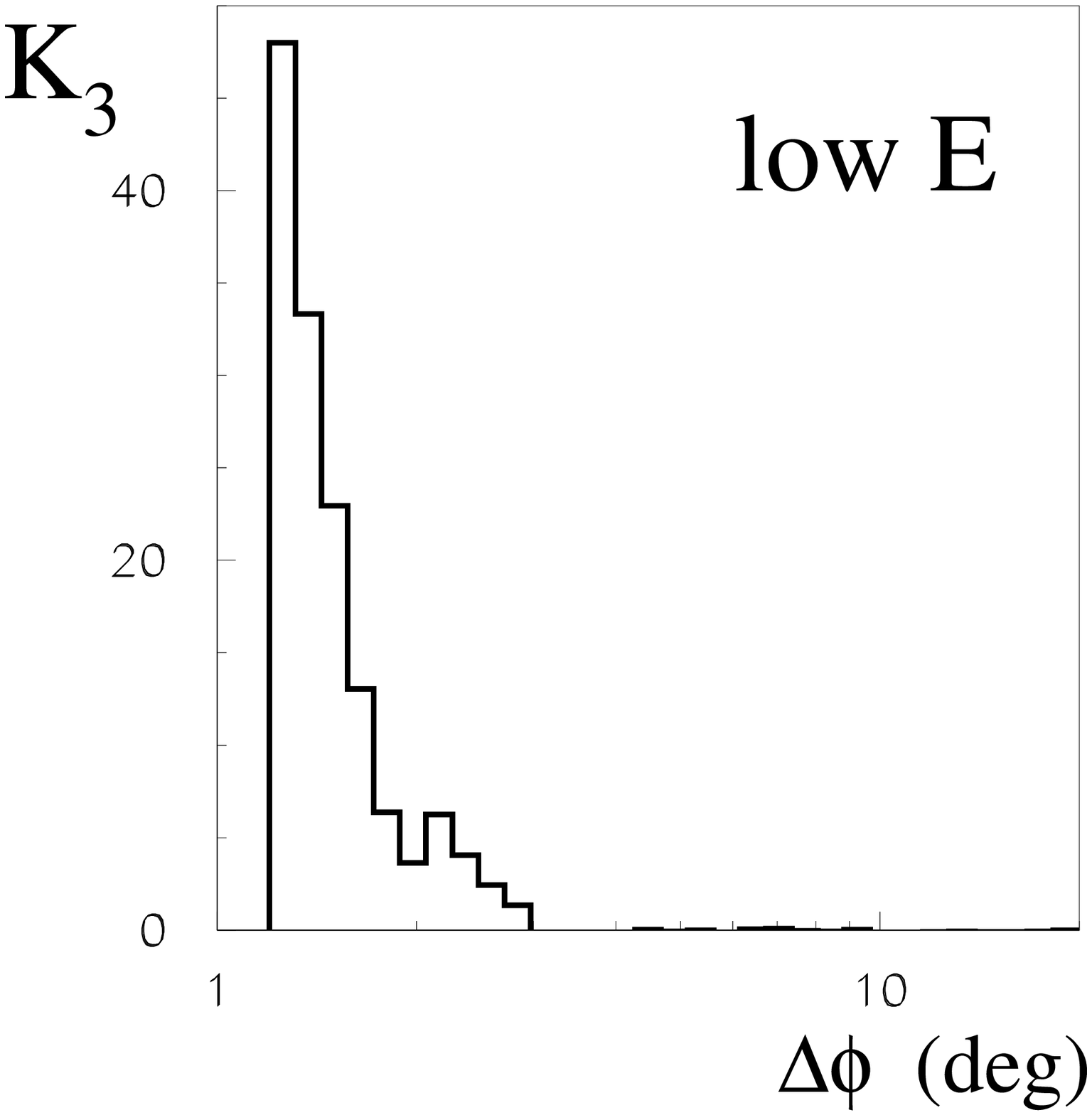,width=7cm}
\psfig{file=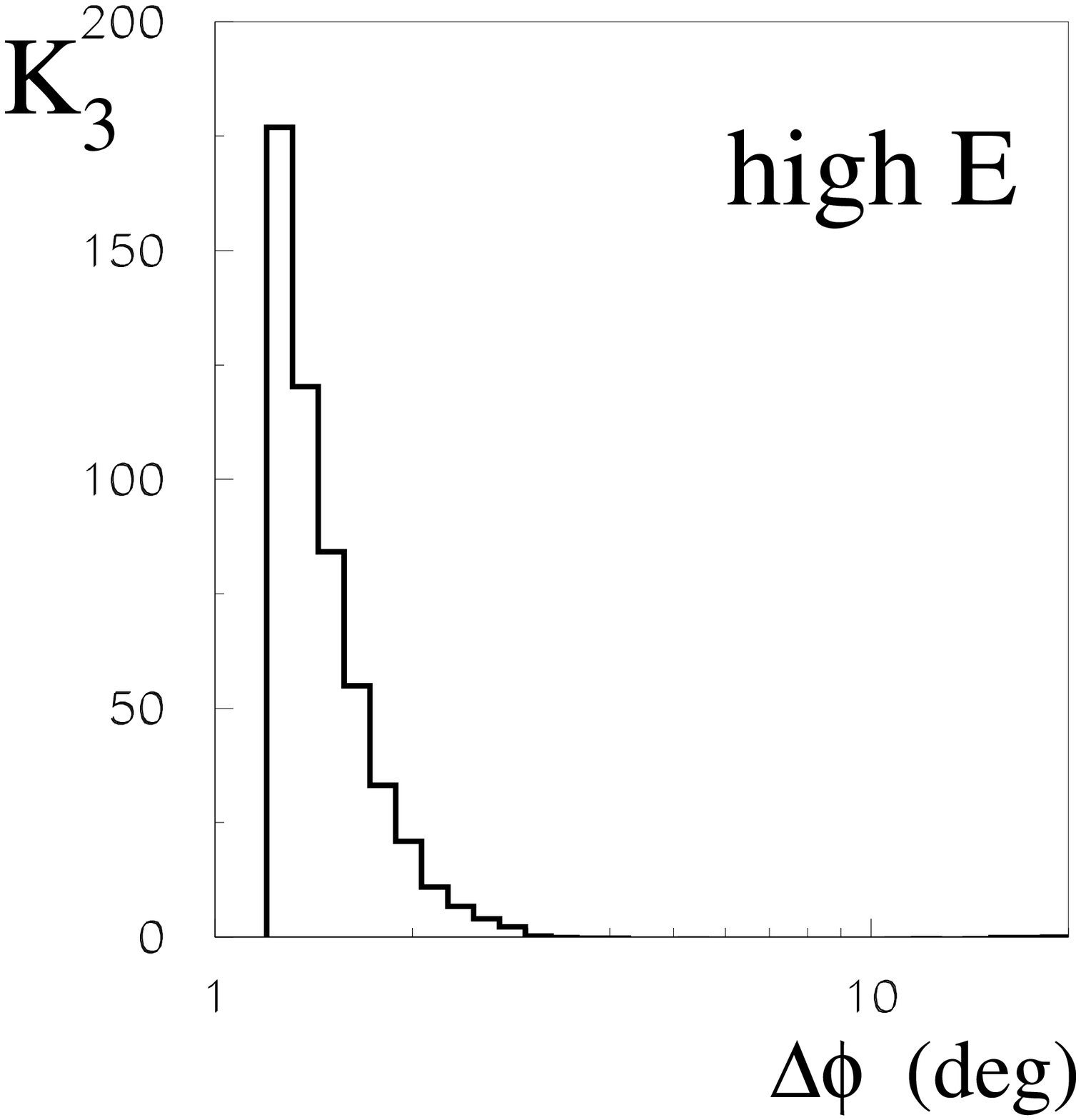,width=7cm}}
 \caption{Third order factorial moments (top) and cumulants (bottom) showing
genuine threefold correlation
calculated for all event sample (left) and the one representing higher energy event
triplets (right).
%and for triplets for which at last one event is of
%energy greater than $10^{20}$~eV (right).
\label{f3k3}}
\end{figure}

Results on third order factorial moments and cumulants are shown in Fig.~\ref{f3k3}.

Concerning doublets analysis ($F_2$),
clustering appearing below 3--4$^\circ$
can be seen
for both data samples. The probability that this is pure coincidence
is of the order of 1\%.
For triplets ($F_3$), the same can be said, suggesting that
very close events may really exist in the data
(but they are still at low confidence level).

It is known \cite{clusagasa} that there is one very close triplet in the data.
Its angular dimension is of the order of experimental angle determination accuracy,
estimated to be a few degrees. There is a possibility that it is a
real cluster
correlated with a UHECR source superimposed
on all the other isotropic UHECR directions.
To generalize this concept (however
the low (1-5) percent confidence is too small for any radical claims),
we can try to find out how strong
the real correlation should be to produce the effect.
The hypothesis in question is that UHECR arrive in most cases from completely
random directions, but there is a small probability that the single UHECR event
is accompanied by another one from the same (within a few degrees) direction.
The factorial moment method allows us to examine this hypothesis in a straightforward
way. Making the ``mixing event sample'' to evaluate the denominator in
Eq.(\ref{fact}), we can make it in a not exactly random way,
but in the way described above, introducing a new parameter---the probability of
accompaniment. With such a constructed
reference sample, the full analysis
can be performed, giving as a result the cumulants
equal to 0 (and $F_2$ equal to 1),
if the real data follow the ``additional accompaniment'' idea.

\begin{figure}[ht]
\centerline{
\psfig{file=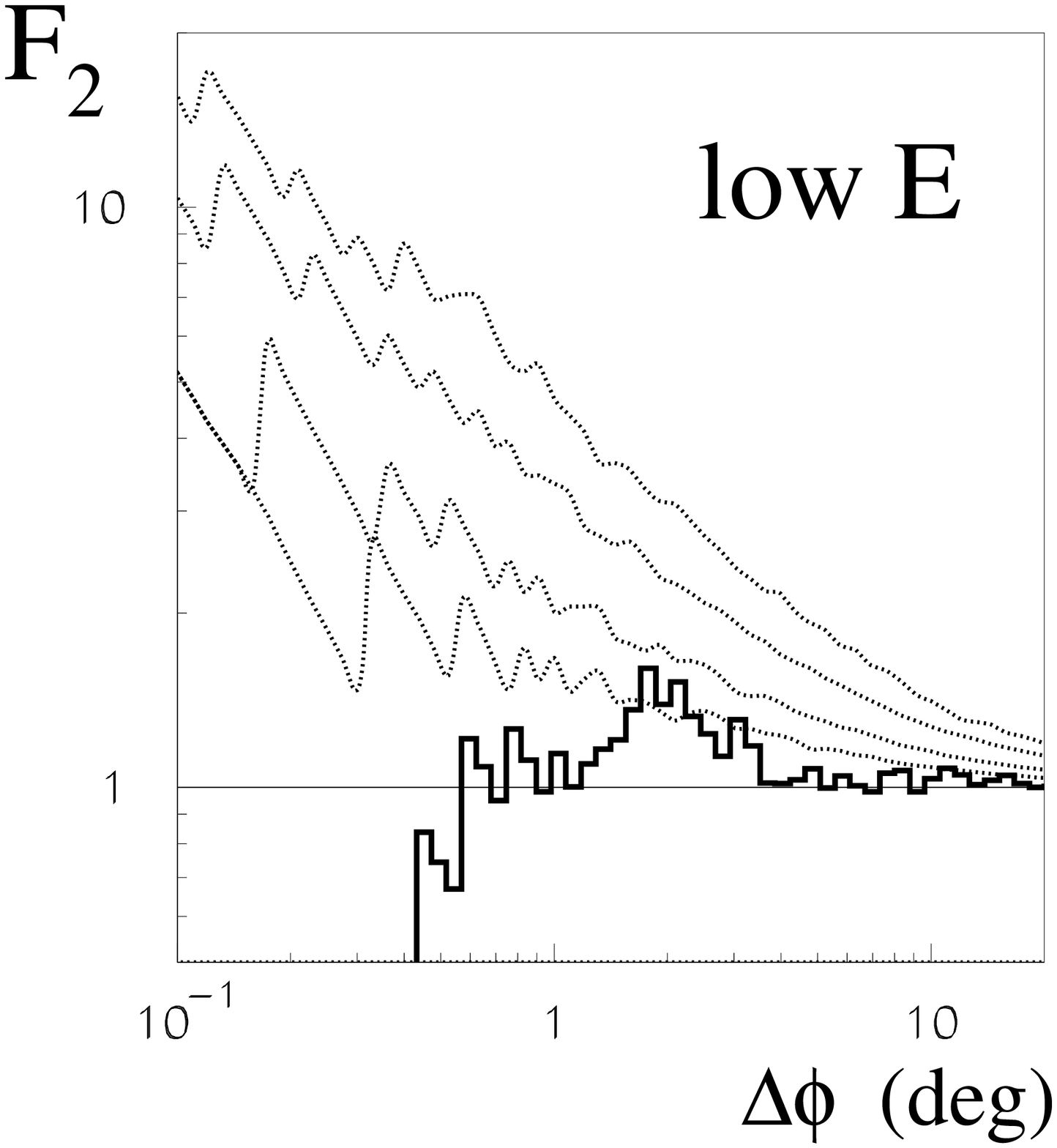,width=7cm}
\psfig{file=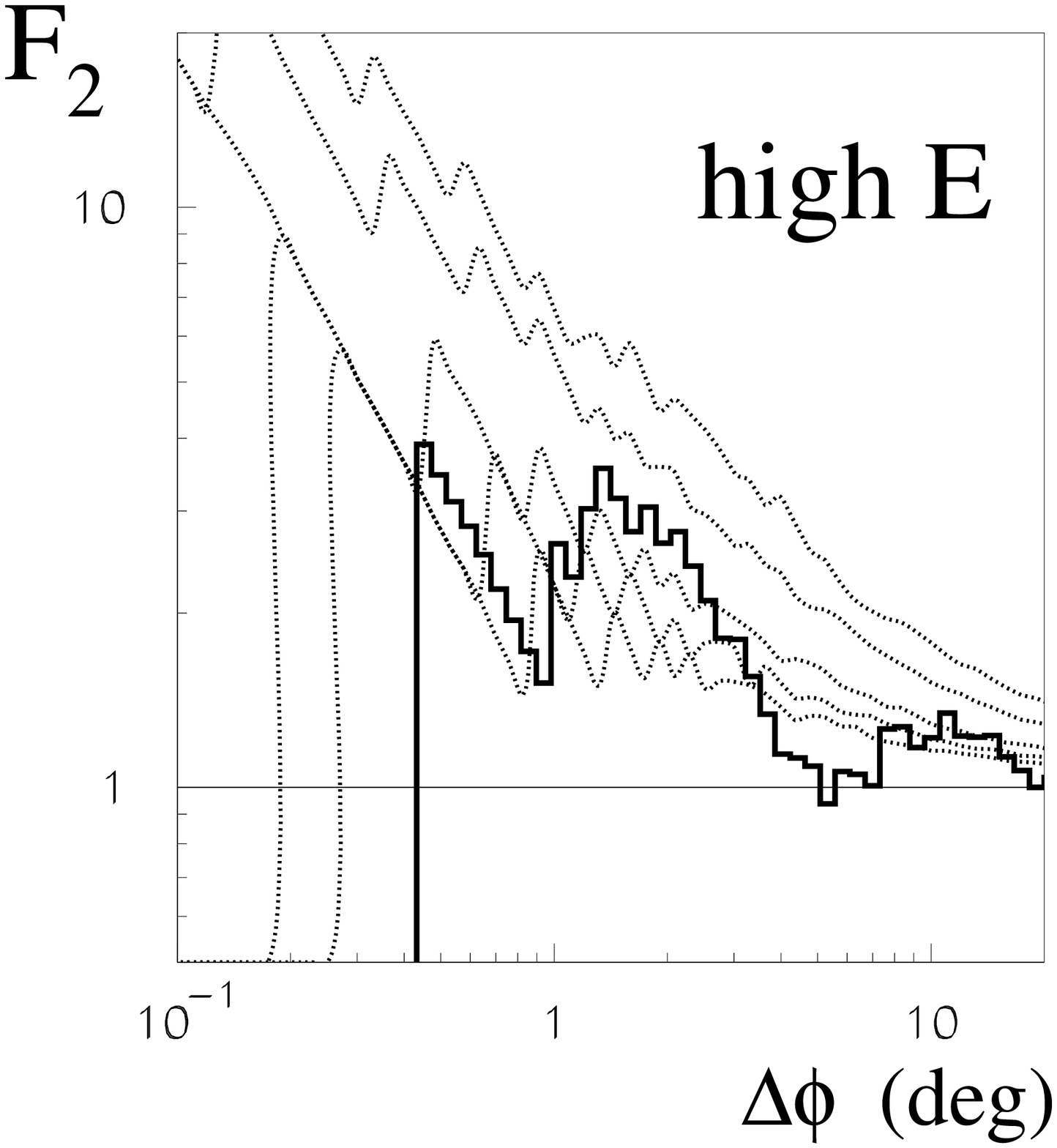,width=7cm}}

\centerline{
\psfig{file=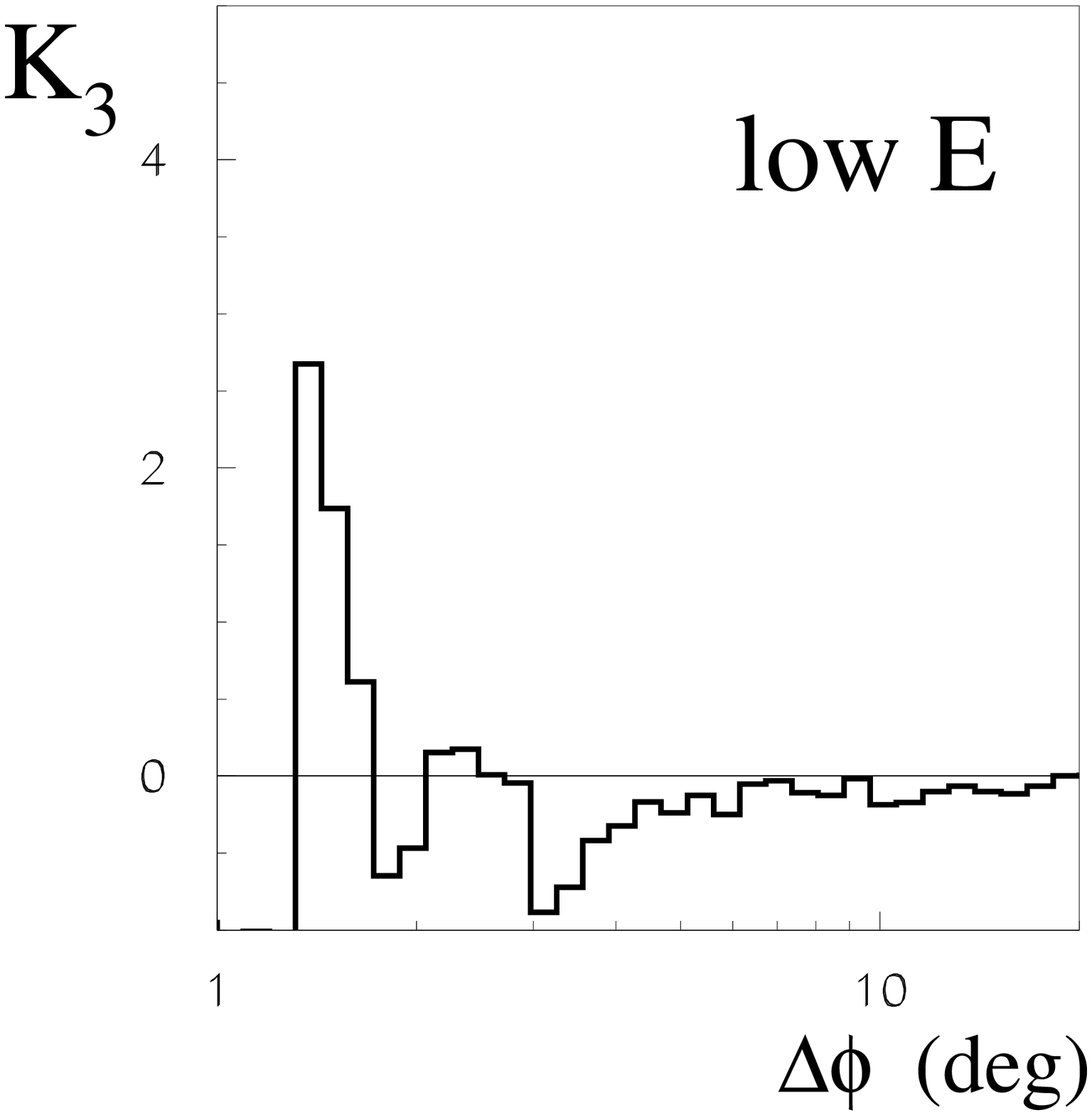,width=7cm}
\psfig{file=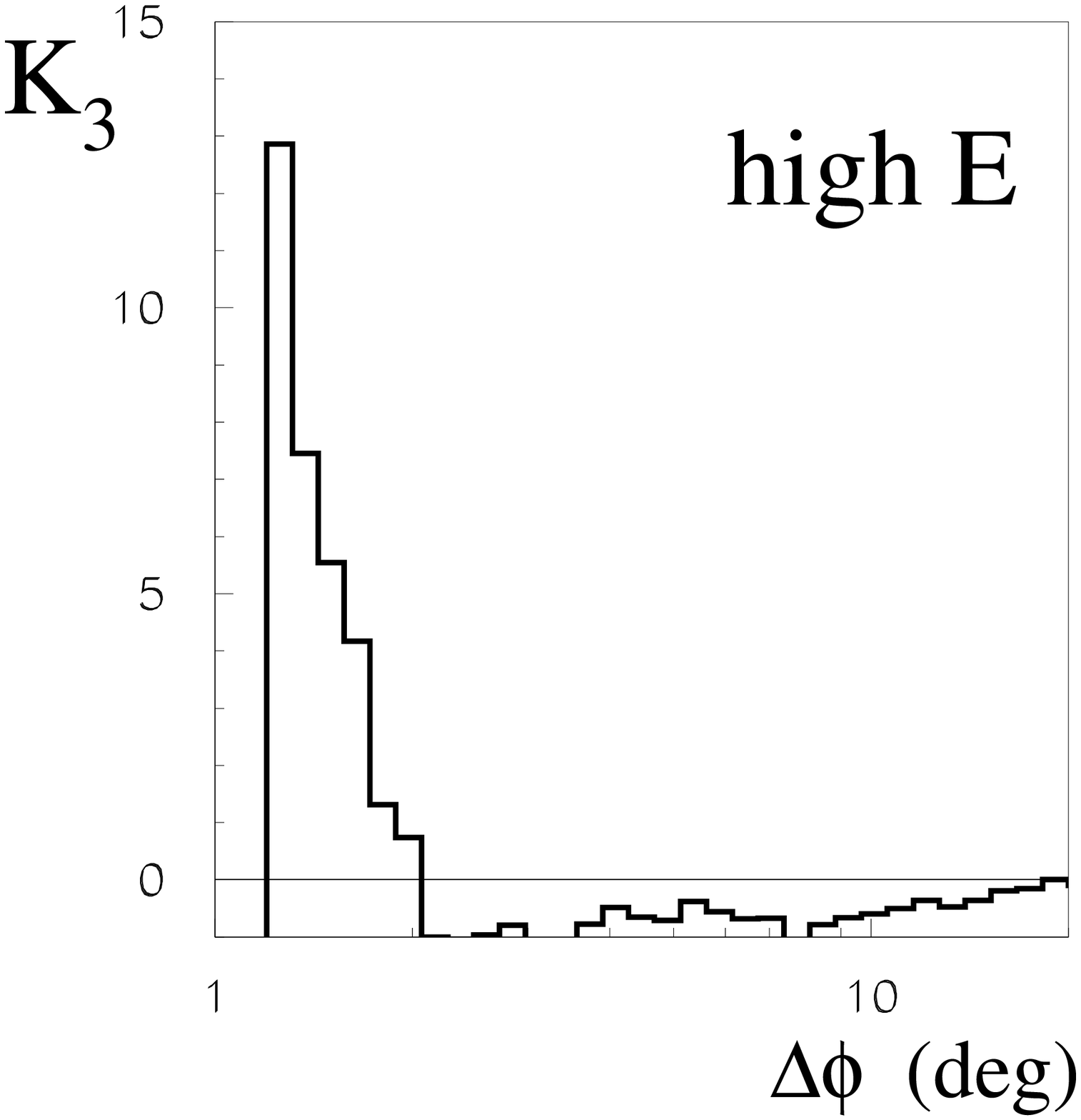,width=7cm}}
\caption{Second order factorial moments and third order cumulants for the model with artificially introduced
correlation (accompaniment probability of 3\%)
calculated for all triplets (``low E'')
and only for those including one UHECR of energy greater than 10$^{20}$~eV (``high E'').
\label{f3k3c}}
\end{figure}

In Fig.~\ref{f3k3c}, we present the results of such an analysis with the additional
accompaniment probability equal to 3\%.
This means that on average, the ``mixed sample'' contains
a few ($\sim$3) close artificial doublets
(there are $113$ events in the sample) for the ``low energy'' case (all events with $E>4\times 10^{19}$~eV).
This is certainly not a big number, but one can see that the
difference it makes is quite substantial.
Summarizing, the existence of close clusters of UHECR, if real (hypothesis verified by existing
data at the 95-99\% confidence level), can be interpreted as the
small, on the level of one percent, probability that there exist a few UHECR sources emitting
particles reaching Earth with directions pointing to the source.
Because we are working with statistics of correlated events in the data
of order of a few, there is nothing more which can be said with reasonable confidence.

\section{Deviation angles}

\subsection {Non-interacting protons}
To examine the small scale clustering (at most on the few percent level, as was shown above), and
to answer the questions: (i) do they create a nuisance for the ``single UHECR source'' model,
and, (ii) in general, what are predictions for the propagation calculations
in realistic intergalactic magnetic fields, and
what deviations can be expected, extensive Monte Carlo
calculations are needed.

For a given particle energy, the angle between the direction to the source and the observed
particle velocity is called the deviation angle.
This angle depends finally on the distance from the source and the time of
particle propagation. Even for particle energies and distances
for which rectilinear propagation dominates, there are
particle trajectory fluctuations which allow the particle
to be observed long after the $R/c$ time and with big
deviation angles. Because we are interested in fractions of events as small
as 3\%, these fluctuations could be important.

To see the effects of the extragalactic magnetic field, the calculations were
first performed for singly charged
particles without any energy loss processes.
In Table \ref{table1}, the fractions of particles arriving in given delay
time intervals are given, and in Table \ref{table2},
average deviation angles are given for different particle energies
and source distances with five ranges of
delay time (with respect to the light signal).
The first delay time range contains UHECR which propagate
almost rectilinearly, and the last range contains
the diffusive component whose velocities are oriented
completely randomly.

\begin{table}[ht]
\begin{center}
\begin{tabular}{r|ccccccccccc}\hline
          { 5 Mpc}&  $5\times10^{17}$&  $10^{18}$&  $2\times$&  $5\times$&  $10^{19}$&  $2\times$&  $5\times$&  $10^{20}$&  $2\times$&  $5\times$&  $10^{21}$\\
\hline\hline
 $\tau<10^{5}$        &      &     &     &     2&    18&    86&   100&   100&   100&   100&   100\\
  $10^{5}<\tau<10^{6}$&      &     &     1&    22&    70&    12&     &     &     &     &     \\
   $10^{6}<\tau<10^{7}$&     &     1&    10&    29&     4&     &     &     &     &     &     \\
   $10^{7}<\tau<10^{8}$&     2&     6&    24&    13&     &     &     &     &     &     &     \\
   $\tau>10^{8}$      &     98&    93&    65&    34&     8&     2&   &     &     &     &     \\
       \hline
\multicolumn{12}{c}{ }\\
\hline { 15 Mpc}  &  $5\times10^{17}$&  $10^{18}$&  $2\times$&  $5\times$&  $10^{19}$&  $2\times$&  $5\times$&  $10^{20}$&  $2\times$&  $5\times$&  $10^{21}$\\
\hline \hline
  $\tau<10^{5}$&     &     &     &     &     &     2&    42&    91&   100&   100&   100\\
  $10^{5}<\tau<10^{6}$&     &     &     &     &     4&    72&    57&     9&     &     &     \\
   $10^{6}<\tau<10^{7}$&     &     &     &     6&    34&    22&     1&     &    &     &     \\
   $10^{7}<\tau<10^{8}$&     &     &     2&    17&    15&     &     &     &     &     &     \\
   $\tau>10^{8}$&   100&   100&    98&    77&    47&     4&     &     &     &   &     \\
        \hline
\multicolumn{12}{c}{ }\\
\hline{ 50 Mpc}  &  $5\times10^{17}$&  $10^{18}$&  $2\times$&  $5\times$&  $10^{19}$&  $2\times$&  $5\times$&  $10^{20}$&  $2\times$&  $5\times$&  $10^{21}$\\
\hline\hline
  $\tau<10^{5}$       &     &     &     &     &     &     &     &     4&    60&    94&   100\\
  $10^{5}<\tau<10^{6}$&     &     &     &     &     &     &    23&    78&    37&     6&     \\
   $10^{6}<\tau<10^{7}$&     &     &     &     &     &    27&    65&    18&     2&     &    \\
   $10^{7}<\tau<10^{8}$&     &     &     &     &     6&    22&    7&     &     &     &     \\
   $\tau>10^{8}$       &   100&   100&   100&   100&    94&    51&     6&    &    &
   &\\
\hline
\end{tabular}
\end{center}
\caption{Fractions of cosmic ray flux
(non-interacting protons)
arriving in given delay time intervals for different
particle energies at different distances to the source.
\label{table1}}
\end{table}

\begin{table}[ht]
\begin{center}
\begin{tabular}{r|ccccccccccc}
\hline
        { 5 Mpc} &  $5\times10^{17}$&  $10^{18}$&  $2\times$&  $5\times$&  $10^{19}$&  $2\times$&  $5\times$&  $10^{20}$&  $2\times$&  $5\times$&  $10^{21}$\\
\hline\hline
  $\tau<10^{5}$       &      &      &      &     9&     6&     5&     3&     1&     1&      $<$1&    $<$1 \\
  $10^{5}<\tau<10^{6}$&      &     &    27&    21&    17&    12&      &      &      &      &     \\
  $10^{6}<\tau<10^{7}$&   &    52&    55&    47&    30&      &      &      &      &      &     \\
  $10^{7}<\tau<10^{8}$&    79&    77&    78&    81&    &      &      &      &      &      &     \\
  $\tau>10^{8}$       &    90&    90&    90&    90&    90&    90&      &      &      &      &     \\
       \hline
\multicolumn{12}{c}{ }\\
       \hline { 15 Mpc}  &  $5\times10^{17}$&  $10^{18}$&  $2\times$&  $5\times$&  $10^{19}$&  $2\times$&  $5\times$&  $10^{20}$&  $2\times$&  $5\times$&  $10^{21}$\\
\hline\hline
  $\tau<10^{5}$       &      &      &      &      &      &     6&     4&     3&     1&     1&     $<$1\\
  $10^{5}<\tau<10^{6}$&      &      &      &      &    12&    11&     8&     8&      &      &     \\
  $10^{6}<\tau<10^{7}$&      &      &      &    37&    30&    23&     25&      &      &      &     \\
  $10^{7}<\tau<10^{8}$&      &    &    68&    69&    73&      &      &      &      &      &     \\
  $\tau>10^{8}$&    89&    90&    90&    90&    90&    90&     &      &      &      &     \\
         \hline
\multicolumn{12}{c}{ }\\
         \hline { 50 Mpc}&  $5\times10^{17}$&  $10^{18}$&  $2\times$&  $5\times$&  $10^{19}$&  $2\times$&  $5\times$&  $10^{20}$&  $2\times$&  $5\times$&  $10^{21}$\\
\hline\hline
  $\tau<10^{5}$       &      &      &      &      &      &      &      &     2&     2&     1&     1\\
  $10^{5}<\tau<10^{6}$&      &      &      &      &      &      &     7&     5&     4&     3&     \\
  $10^{6}<\tau<10^{7}$&      &      &      &      &    &    20&    15&    13&     10&      &     \\
  $10^{7}<\tau<10^{8}$&      &      &      &    &    38&    40&    40&    &      &      &     \\
  $\tau>10^{8}$&    89&    90&    90&    90&    90&    90&    90&      &      &
  &\\
  \hline
\end{tabular}
\end{center}
\caption{Mean deviation angle for non-interacting protons  propagating from a
source located at different distances
as a function of particle energy and delay time. For some particular values,
the flux of particles is negligible,
so the values cannot be given.
\label{table2}}
\end{table}

It can be easily seen that for singly charged
particles of energies of about $5\times10^{19}$~eV,
if the source is within
$\approx$ 15 Mpc, the propagation
is almost rectilinear, delay times are not bigger than
10$^6$~years, and mean deviation angles are
less than 10$^\circ$. For particles of energy greater
than 10$^{20}$~eV, the mean deviation is very small, even if the
particles come from 50 Mpc away.

The propagation of non-interacting particles scales with $Z$.
To see what the situation is with iron nuclei, one
has to look in Tables~\ref{table1} and \ref{table2} for energies
26 times smaller.
The iron nucleus of energy $5\times10^{19}$~eV
travels on average longer than 10$^8$~years and arrives almost isotropically
even for sources as close as 5 Mpc. For a source at 15 Mpc and
energy of 10$^{20}$~eV, still no trace of anisotropy can be
expected.

This situation, however, can change if energy
loss processes are taken into account.
Particles traversing intergalactic space
can interact with the matter and fields there.
The longer they propagate, the bigger
the energy losses are. It is expected that the
general effect of UHECR interactions will be to favor the shorter
paths, corresponding to smaller delays and deviation angles.

\subsection {Introduction of energy loss processes}

Results of calculations for the propagation of protons and iron
nuclei are shown in Tables~\ref{table3} and \ref{table4}.

\begin{table}[ht]
\begin{center}
\begin{tabular}{r|ccccccccccc}
\hline
        protons  &  $5\times10^{17}$&  $10^{18}$&  $2\times$&  $5\times$&  $10^{19}$&  $2\times$&  $5\times$&  $10^{20}$&  $2\times$&  $5\times$&  $10^{21}$\\
\hline\hline
          $\tau<10^{5}$&            &     &     &     &     &    2&   29&   98&  100&  100&  100 \\
          $10^{5}<\tau<10^{6}$&     &     &     &     &    2&   71&   71&    2&     &     &     \\
          $10^{6}<\tau<10^{7}$&     &     &     &    3&   79&   27&     &     &     &     &     \\
          $10^{7}<\tau<10^{8}$&     &     &     &   25&   10&     &     &     &     &     &     \\
          $\tau>10^{8}$&         100&  100&  100&   72&    9&     &     &     &     &     &     \\
       \hline
\multicolumn{12}{c}{ }\\
       \hline iron  &  $5\times10^{17}$&  $10^{18}$&  $2\times$&  $5\times$&  $10^{19}$&  $2\times$&  $5\times$&  $10^{20}$&  $2\times$&  $5\times$&  $10^{21}$\\
\hline\hline
          $\tau<10^{5}$       &     &     &     &     &     &     &     &     &     &        &         \\
          $10^{5}<\tau<10^{6}$&     &     &     &     &     &     &     &     &   20&        &         \\
          $10^{6}<\tau<10^{7}$&     &     &     &     &     &     &    1&   18&   61&        &         \\
          $10^{7}<\tau<10^{8}$&     &     &     &     &     &     &   11&   49&   19&        &         \\
          $\tau>10^{8}$       &  100&  100&  100&  100&  100&  100&   88&   33&     &        &         \\
          \hline
\end{tabular}
\end{center}
\caption{Fractions of cosmic ray UHECR flux arriving in given delay time intervals for different (observed)
(observed) energies. Source distance is 15 Mpc.\label{table3}}
\end{table}

\begin{table}[ht]
\begin{center}
\begin{tabular}{r|ccccccccccc}
\hline
     protons &  $5\times10^{17}$&  $10^{18}$&  $2\times$&  $5\times$&  $10^{19}$&  $2\times$&  $5\times$&  $10^{20}$&  $2\times$&  $5\times$&  $10^{21}$\\
\hline\hline
          $\tau<10^{5}$       &      &      &      &       &      2&      2&      4&      3&      1&     1&      $<$1  \\
          $10^{5}<\tau<10^{6}$&      &      &      &      4&     12&     11&      7&      5&       &      &        \\
          $10^{6}<\tau<10^{7}$&      &      &    20&     39&     28&     18&     20&       &       &      &        \\
          $10^{7}<\tau<10^{8}$&      &      &    62&     57&     59&       &       &       &       &      &        \\
          $\tau>10^{8}$       &    90&    90&    90&     90&     90&       &       &       &       &      &        \\
       \hline
\multicolumn{12}{c}{ }\\
       \hline { iron}  &  $5\times10^{17}$&  $10^{18}$&  $2\times$&  $5\times$&  $10^{19}$&  $2\times$&  $5\times$&  $10^{20}$&  $2\times$&  $5\times$&  $10^{21}$\\
\hline\hline
          $\tau<10^{5}$       &       &       &       &       &       &       &       &       &       &      &        \\
          $10^{5}<\tau<10^{6}$&       &       &       &      2&      9&      3&      6&      3&      2&      &        \\
          $10^{6}<\tau<10^{7}$&       &     58&     19&     20&     15&       &     24&     24&     29&      &        \\
          $10^{7}<\tau<10^{8}$&       &     41&     39&     49&       &     33&     43&     62&     69&      &        \\
          $\tau>10^{8}$       &    90&     90&     90&     90&     90&     90&     90&     90&        &      &        \\
\hline
\end{tabular}
\end{center}
\caption{Mean deviation angle for UHECR  from the source located
at 15 Mpc as a function of particle energy and delay time.\label{table4}}
\end{table}

From Tables~\ref{table3} and \ref{table4}, for about 10\% of the events, the mean
deviation for iron nuclei is about
40$^\circ$ at an energy of $5\times10^{19}$~eV. Going a little further up in
particle energy, to 10$^{20}$~eV, approximately 20\% have mean
deviation angle of about 20$^\circ$ (delays less than 10$^7$~years)
and the next 50\%
have mean deviation 40$^\circ$ and arrive not later than 10$^8$~years
after the light signal.
This is enough to see some slight enhancement of UHECR from the
region on the sky where the source is (where galaxies collide? \cite{alda}), but obviously the
general anisotropy constraint is still
fulfilled. It is easy to achieve
more or less anisotropy because there is some freedom with the magnetic
fields (they can be eventually smaller or larger than assumed in this work).

In the case of protons, however, if one wants to see them above 10$^{19}$~eV,
a strong anisotropy has to be observed and the source must be
active at present, so there is the general possibility of the identification
of the UHECR source with some
astrophysical object on the sky.

Concerning the small scale clustering problem, we have to say that between our
results and the widely discussed in the
literature existence vs. lack of coincidences between UHECR direction
and astrophysical objects, two solutions are possible. The first is that the clustering is by pure chance coincidence, which can be
accepted at the 95 or 99\% confidence level. The second possibility is that
there is only one relatively close source of UHECR active ``at present'' (or at least only a few sources)
and protons from there form the cluster(s) in question. In this case, however, the bulk
of UHECR events are produced in other sources, or in one ``single source'' which is
``at present'' not active.

A review of the actual experimental situation was presented in the XXVIII$^{\rm th}$ International
Cosmic Ray Conference in Tsukuba \cite{icrcclus}.
Recently, the AGASA group confirmed their findings of close doublets and triplets
for energies from $10^{19}$~eV to the very end of the spectrum, and the HiRes experiment
does not see anything like this above $10^{19}$~eV. The problem with high statistic monocular
HiRes data is that the angular resolution is rather poor (and obviously not symmetrical). With such
conditions, however, they give an upper limit for doublets and it is equal to 4.
The more precise HiRes
stereo data set consists only of 164 events above $10^{19}$~eV, and nothing more significant than
1$\sigma$ is seen there.
Both HiRes statements, in spite of being negative, do not contradict
at a high significance level the AGASA statement. On the other hand (as was shown above), the
significance of AGASA clustering is in fact only on the 95-99\% confidence level.

\section{Predicted UHECR flux}
\begin{figure}[ht]
\centerline{\psfig{file=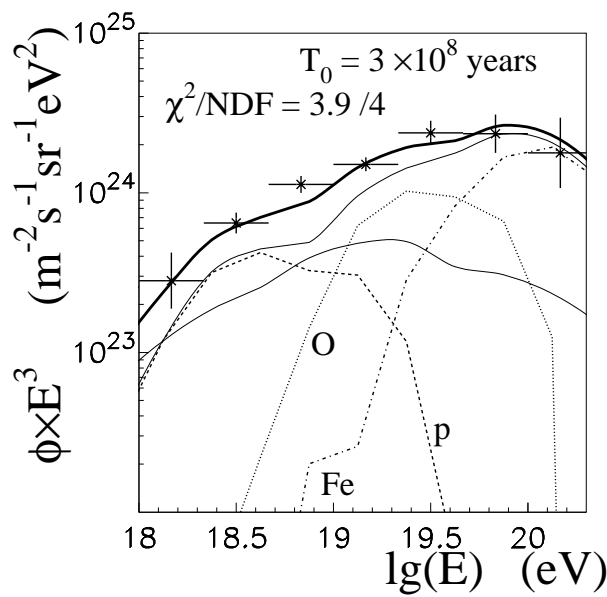,width=8.5cm}}
\caption{UHECR flux for Single Source.
Source is located at 15 Mpc and switched off at
the time $T_0=3\times10^8$~years ago.  The proton contribution is shown by dashed,
iron by dot-dashed, and oxygen by
dotted line. 
%The source power realised in the cosmic rays
%is in all three components---
%(10:5:3)$\times 10^{59}$~erg, respectively, 
%for the whole time of source activity (=~$10^{8}$~years).
The source spectrum index is $-2.1$.
The continuous
background is shown by the thin solid line and was obtained assuming one random source
per each 1000~Mpc$^3 \times 10^9$~years.
\label{spes}}
\end{figure}

The exact propagation calculations were performed for
different times of source activity. The source composition of
protons and iron and oxygen nuclei was assumed, and their relative
proportions were adjusted to the
data on the extragalactic UHECR flux \cite{sww}. The continuous background consisting of
cosmic rays produced by sources identical to the Single Source,
uniformly distributed in the Universe (one per 1000~Mpc$^3$ per
10$^9$ years) was also assumed. Its contribution is
given in Fig.~\ref{spes} by the thin solid line. This background
is about 10\% of the total UHECR flux and does not play
a significant role here.

The experimental points used in the present work
are taken from the analysis of the Northern hemisphere ``world data''
given in Ref.\cite{sww}. The points shown in Fig.\ref{spes} represent the combined UHECR spectrum
from Haverah Park, AGASA, Volcano Ranch, Yakutsk and Fly's Eye experiments, after
subtraction of the galactic
component (dominating still at about 10$^{18}$~eV, but negligible above 10$^{19}$~eV). The agreement between
different measured data sets was achieved by adjusting the individual energy estimation accuracy and
overall normalization.  It was found satisfactory, thus allowing the authors to give the extragalactic UHECR energy
spectrum free of particular instrumental biases.

The very recent discussion of the UHECR spectrum \cite{icrcspec}
and the discrepancies between the last reported AGASA spectrum which
contains 11 events of energies above $10^{20}$~eV
and HiRes (mono and stereo) spectra (compatibile with the Fly's Eye spectrum)
with only 2 such cases with comparable exposures
shows that there is, probably, systematic bias in energy determination on the order of 30\% in one of the
experiments.
The method of combining energy spectra applied in Ref.\cite{sww} takes into account such
uncertainties. Thus, experimental points in Fig.\ref{spes} represent well the actual situation
in the UHE region. The problem of whether the GZK cut-off exists (as claimed by the AGASA group) or not
(according to HiRes) is still an open question, but the data seems to be as shown
in the figure.

The number of parameters adjusted to the seven points representing extragalactic UHECR flux
at first sight seems to be unreasonably big: normalization, composition (2 parameters), source spectral index,
$T_0$, source activity duration time, distance to the source,
background normalization (density of the sources averaged in large space and time intervals)---
all together eight of them. This is big if all of the parameters are uncorrelated. But they are in fact
strongly correlated, and the freedom of choice does not represent the real number of degrees of freedom
of the fitting procedure.

It is clear that the ``best'' spectrum shown in Fig.\ref{spes}
does not match exactly the experimental points used.
Additionally, it can be mentioned here that some of these parameters
are fixed to some reasonable boundaries (as spectral index or background normalization). In general the main
purpose of the present paper is not to fit the data perfectly by a single line;
the uncertainties of many
astrophysical parameters of extragalactic space (magnetic field structure, matter and radiation densities, etc.)
and the limited statistics of registered UHECR events do not permit the derivation of
any strong physical conclusions from this kind of fit. We want, rather, to show only that general
agreement (or, more precisely, lack of experimental contradiction) can be achieved
within the proposed UHECR origin model.
Thus the particular values of $T_0$ , p:O:Fe composition, and distance to the source
taken to draw the lines in Fig.\ref{spes} should be treated not as the main
result of this work, but rather as an
example, showing that with values like these,
quite satisfactory agreement between the ``single UHECR source model''
and the measured extragalactic UHECR spectrum can be obtained.

Our fit to the UHECR spectrum was found with the
Single Source inactive for the last $3\times 10^8$~years. The composition
(p:O:Fe about 10:5:3)
is
%--------------------------------------------------------typical.
not extraordinary if compared with the one derived from
experimental information in the
energy region of about three orders of magnitude lower, {\it i.e.,} at ``the knee.''
%%--------------------------------------------------------typical.
It is important to mention that for such ``light''
source composition, the observed UHECR flux above
$3\times10^{19}$~eV is quite heavy, and above $10^{20}$~eV
is completely iron-dominated.

More detailed discussion of the particular parameter values obtained is given in \cite{ww2004}.

\section{Conclusions}

Extensive Monte Carlo calculations for the propagation of UHECR in
extragalactic magnetic fields have been performed.

The directional small scale clustering in the available data gives a
5\% limit on its chance origin. The data are consistent
with the hypothesis that the overall isotropic UHECR direction distribution is enhanced by
the additional clustering probability on the level of no more than a
few ($\sim3$) percent of observed UHECR events.
Propagation calculations show that such enhancement can be related
to primary protons from only the source (or sources) active at present.

It is possible that the bulk of the UHECR are created in a Single Source
at 15 Mpc distance which was active about $3\times 10^8$~years ago.
This model is consistent with the large scale
anisotropy data (most of the flux is composed of isotropized iron and heavy nuclei),
as well as with the measured flux of extragalactic cosmic rays
of energies above 10$^{18}$~eV.
No new physics concerning the GZK cut-off mechanism is needed.

%%%%%%%%%%%%%%%%%%%%%%%%%%%%%%%%%%%%%%%%%%%%%%%%%%%%%%%%%%%%%%%%%%%%%%%%%%%%%%%%%%%%%%%%%%%%%%%%%%%%%%%%%%%%%%%%%
\section*{Acknowledgment}
The author is grateful to Arnold Wolfendale for valuable discussions and helpful
comments.

\end{document}